\documentclass{article}

\usepackage{geometry}
\usepackage{amssymb,amsbsy,amsmath,amsfonts,amscd}
\usepackage{latexsym,euscript,exscale,epic,eepic,epsfig}
\usepackage{float}
\usepackage{calc}
\usepackage{bm}
\usepackage{graphicx}
\usepackage{enumerate}
\usepackage{color}
\usepackage{pstricks,pst-node,pst-text,pst-3d,pst-plot}
\usepackage[subnum]{cases}
\usepackage[percent]{overpic}
\usepackage{subfig}
\usepackage{graphicx}

\def \tr{\mathop{\rm tr }\nolimits}

\DeclareMathOperator*{\argmin}{arg\,min}
\def\doubleunderline#1{\underline{\underline{#1}}}

\usepackage[utf8]{inputenc}
\usepackage{verbatim}
\usepackage{listings}
\usepackage{algorithm}
\usepackage{mathtools}
\usepackage{subfig}
\usepackage{multirow}
\usepackage{enumitem}

\usepackage{authblk}
\usepackage[numbers,sort&compress,merge]{natbib}
\usepackage{hyperref}
\hypersetup{colorlinks=true, citecolor=blue}
\usepackage{xcolor}

\title{Variational phase-field continuum model uncovers adhesive wear mechanisms in asperity junctions}

\author[1]{Sylvain Collet}
\author[1]{Jean-François Molinari}
\author[2]{Stella Brach\footnote{Corresponding author: stella.brach@polytechnique.edu}}
\affil[1]{{\small Institute of Civil Engineering, Institute of Materials Science and Engineering, Ecole Polytechnique Fédérale de Lausanne (EPFL), CH 1015 Lausanne, Switzerland}}
\affil[2]{{\small Laboratoire de Mécanique des Solides, CNRS, École Polytechnique, Institute Polytechnique de Paris, Palaiseau, 91128, France}}
\date{}
\begin{document}
\maketitle

\begin{abstract}
Wear is well known for causing material loss in a sliding interface. Available macroscopic approaches are bound to empirical fitting parameters, which range several orders of magnitude. Major advances in tribology have recently been achieved via Molecular Dynamics, although its use is strongly limited by computational cost. Here, we propose a study of the physical processes that lead to wear at the scale of the surface roughness, where adhesive junctions are formed between the asperities on the surface of the materials. Using a brittle formulation of the variational phase-field approach to fracture, we demonstrate that the failure mechanisms of an adhesive junction can be linked to its geometry. By imposing specific couplings between the damage and the elastic energy, we further investigate the triggering processes underlying each failure mechanism. We show that a large debris formation is mostly triggered by tensile stresses while shear stresses lead to small or no particle formation. We also study groups of junctions and discuss how microcontact interactions can be favored in some geometries to form macro-particles. This leads us to propose a classification in terms of macroscopic wear rate. Although based on a continuum approach, our phase-field calculations are able to effectively capture the failure of adhesive junctions, as observed through discrete Molecular Dynamics simulations.

\end{abstract}

\section{Introduction}
\label{sec:introduction}

Wear is a major aging process of any system experiencing relative motion of its components.  The presence of adhesive forces at the level of the surface roughness, leading to the detachment of wear particles, was long observed and described as adhesive wear \cite{holm-1946,rabinowicz-1958,rabinowicz-1966, rabinowicz-1966}.  The size of the wear particles (besides having a great implication on the global wear rate and thus durability) is strongly related to the resulting pollution. For instance, it was measured that 20\% of the overall traffic-related fine particle emissions comes from brake wear \cite{grigoratos-2015}. These airborne particles are known to cause health hazards, whose intensity is strongly related to the particle size, as nanoscale fragments can leach and settle into internal organs \cite{samet-2000,pope-2002,olofsson-2011}.  One of the first but still most used wear formula, known as the Archard wear law, linearly relates the resulting wear volume to the normal load \cite{archard-1953}. However, this linear relation is only valid for a certain range of applied load \cite{archard-1956} and strongly depends on a fitting parameter, the wear coefficient, that can range over several orders of magnitude \cite{meng-1995}. In presence of lower or higher applied load, the relation between wear volume and applied load was observed to be sublinear or superlinear respectively \cite{kitsunai-1990,hokkirigawa-1991,wang-1996,hsu-2004,kato-2002}. As a consequence, a wear process has to be classified into low, mild or sever wear for an approximate measure of the wear volume to be computed.

Major advances in tribology were recently made possible by the use of Molecular Dynamics (MD), where the wear process can be investigated directly at the level of the surface roughness \cite{aghababaei-2016,aghababaei-2017,aghababaei-2018,brink-2019,milanese-2019}.
In  \cite{aghababaei-2016}, a unified approach was  proposed to describe both the plastic flattening of the surface roughness associated to the low wear regime, and the fracture-induced particle formation associated to the mild wear regime. Authors postulated the existence of a critical length-scale $d^*$, describing the transition from one regime to the other based on the size of the adhesive junction: their computations showed that junctions smaller than $d^*$ were characterized by the gradual smoothing of the asperities, while for sizes larger than $d^*$ a wear debris was detached from the interface by propagation of two cracks. This model was further developed in \cite{brink-2019} to account for weaker adhesion forces at the junction. This new formulation allowed to capture a third mode of failure: the relative slip of the asperities, where damage is only located along the junction's vicinity. Finally in \cite{aghababaei-2018}, the transition from mild to severe wear was traced back to the stress field interaction at the contact points of an adhesive junction, providing the interesting result in that the size of the wear debris can be deduced from the observed early-stage crack patterns.

Although being a powerful tool for modeling wear, Molecular Dynamics carries its own limitations.  
First, the computational cost of MD simulations limits its application to systems at the nanoscale. Second, as the domain greatly evolves with time, a thorough study of the effect of the junction's geometry on the wear process is practically out of reach. Hence the need of a more agile numerical formulation.

In this paper, we investigate adhesive wear by using a phase-field model of brittle fracture \cite{bourdin-2000, bourdin-2008}, which numerically implements the variational gradient damage formulation \cite{francfort-1998}.
Widely used to predict crack nucleation and propagation in a variety of materials and geometries \cite{bourdin-2014,tanne-2018}, this approach does not require any a-priori assumption on the crack path and the fracture process is independent from the numerical discretization, as long as the mesh size is set smaller than a regularization parameter.
With respect to MD simulations, the variational phase-field model allows us to perform a thorough study of various junctions' geometries at a relatively low computational cost. It further permits to model much larger systems involving materials that are defined by their macroscopic properties (such as Young's modulus and fracture toughness),  thus not requiring any atomic potential characterization. Notwithstanding those key advantages, very few works can however be found in literature on the topic of wear. Particularly noteworthy is the recent contribution \cite{carollo-2019}, where a phase-field fracture model was used to simulate crack patterns in adhesive junctions: results suggested a dependence of the wear process on the geometry of the junction, although the performed simulations did not cover a large enough parameter space to provide exhaustive conclusions. 

This is further investigated in the present paper, whose main contribution is twofold. First, we clearly establish a relation between the failure patterns of the adhesive junctions and their geometries. Second, we propose a classification in terms of macroscopic wear rate, qualitatively describing the material loss of the junction. We are thus able to highlight key differences in the failure of adhesive junctions and provide a new understanding on the various failure mechanisms, which can be associated to different wear particle sizes.

The outline of the paper is as follows. In Section \ref{sec:comp-met} we first recall the formulation of the variational phase-field approach to brittle fracture used in this study. The results of the simulations are then presented in Section \ref{sec:single-junctions}, by considering perfectly-adhesive single junctions. The interaction between multiple asperities is investigated in Section \ref{sec:interactions}. Conclusions are drawn in Section \ref{sec:discussion}. The following notation is used throughout the paper: underlined and double-underlined symbols denote vectors and second-order tensors, blackboard letters indicate fourth-order tensors with the exception of $\mathbb{R}$ denoting the real line, the symbol ``:'' is the double-dot product operator.

\section{Numerical methods}
\label{sec:comp-met}

\subsection{Variational phase-field approach to fracture}
\label{sec:pf}

Consider a domain  $\Omega$ comprised of a linear-elastic isotropic material with stiffness tensor $\mathbb{C}$ (with bulk modulus $K$ and shear modulus $\mu$) and  critical energy-release rate (toughness) $G_\text{c}$. 
We investigate adhesive wear via a variational phase-field model of fracture \cite{francfort-1998, bourdin-2000, bourdin-2008}. A regularized energy function $\mathcal{E}_\ell$ is introduced, expressed in terms of the characteristic length $\ell>0$ and of the damage variable $\alpha \in [0,1]$. The case $\alpha=0$ corresponds to the intact material, whereas $\alpha=1$ to complete fracture. The fracture problem is solved by minimizing $\mathcal{E}_\ell$ with respect to the displacement field $\underline{u}$ and the damage variable $\alpha$
\begin{equation}
  \label{eq:reg-min}
  \left(\underline{u}^\ast,\alpha^\ast\right)\,\,=\,\,\argmin_{\substack{\underline{u}\in\mathcal{K}_\text{u}\\ \dot{\alpha}\geq 0}} \,\mathcal{E}_\ell(\underline{u},\alpha)
\end{equation} 
where $\mathcal{K}_{\text{u}}$ is the set of admissible displacement fields with regard to the applied boundary conditions.

In the original developments \cite{bourdin-2000, bourdin-2008}, the regularized energy function $\mathcal{E}_\ell$ was expressed as 
\begin{align}
  \label{eq:noUni}
  \begin{split}    &\mathcal{E}_\ell(\underline{u},\alpha):= \int_{\Omega} \left(W^\text{el} + W^\text{fr}\right) \,\, d\Omega \\
    & W^\text{el} = \frac{1}{2}\,\,\doubleunderline{\epsilon} : g(\alpha)\mathbb{C} : \doubleunderline{\epsilon} \\
    &  W^\text{fr} = \frac{3G_\text{c}}{8}\left(\frac{\alpha}{\ell}+\ell |\nabla\alpha |^2 \right)
  \end{split}   
\end{align}
where $\doubleunderline{\epsilon} = (\nabla\underline{u}+\nabla\underline{u}^\text{t})/2$ is the strain tensor, $g(\alpha)=\eta+(1-\alpha)^2$ is a degradation function describing the decrease in elastic energy as damage progresses and $\eta$ is a small residual stiffness. 
The energy $\mathcal{E}_\ell$ in Eq.\,\eqref{eq:noUni} is symmetric in tension and in compression, thus material interpenetration can occur under compressive loadings.

To avoid this problem, we adopt the approach proposed by Amor \textit{et al.} \cite{amor-2009} (see also \cite{lancioni-2009} for shear fracture), where the strain energy $W^\text{el}$ is split in spherical and deviatoric parts and depends on the sign of the volume change. Accordingly, we decompose the hydrostatic strain in positive (i.e., $\tr\doubleunderline{\epsilon}^+=\max(0,\tr\doubleunderline{\epsilon})$) and negative (i.e., $\tr\doubleunderline{\epsilon}^-=\tr\doubleunderline{\epsilon}-\tr\doubleunderline{\epsilon}^+$) parts and introduce the following regularized models with contact conditions.
\begin{itemize}
    \item \textbf{Positive-hydrostatic} (PH). Aiming to reproduce mode-I cracks, fracture is only allowed in  material regions with positive volume change. As such, any increase in the fracture energy $W^\text{fr}$ is exclusively due to a reduction in the spherical part of the tensile strain energy
    \begin{equation}
  \label{eq:posH}
    W^\text{el} = \frac{K}{2} \left(\tr^2\doubleunderline{\epsilon}^- +g(\alpha) \tr^2\doubleunderline{\epsilon}^+ \right)+ \mu\left(\doubleunderline{\epsilon}_\text{d}:\doubleunderline{\epsilon}_\text{d}\right)
\end{equation}
where $\doubleunderline{\epsilon}_\text{d}$ is the strain deviator. Conversely, compressive volume changes and shear states do not contribute to the fracture process.
    \item \textbf{Hydrostatic-deviatoric} (HD). With respect to the previous model, mode-II cracks are reproduced by adding the deviatoric part of the strain energy to the coupling with the fracture energy.  The resulting regularized expression for the elastic energy is
\begin{equation}
  \label{eq:hydroD}
  W^\text{el} = \frac{K}{2}\tr^2\doubleunderline{\epsilon}^- +g(\alpha)\left( \frac{K}{2} \tr^2\doubleunderline{\epsilon}^+ + \mu\left(\doubleunderline{\epsilon}_\text{d}:\doubleunderline{\epsilon}_\text{d}\right)\right)  
\end{equation}
where only material regions with positive volume changes or under shear actions are allowed to release strain energy through fracture.
\end{itemize}

For our computations, we refer to a non-dimensional form of the previous models. The regularized functional $\mathcal{E}_\ell$ is divided by $E_0L_0^3$, where $E_0$ is a typical value of the Young's modulus and $L_0$ is a characteristic length of the domain. The resulting non-dimensional energy function $\widetilde{\mathcal{E}}_\ell$ reads as in Eq.\,\eqref{eq:noUni}, except that each quantity is now replaced by its non-dimensional counterpart
\begin{equation}
  \label{eq:no-dim}
    \widetilde{\mathbb{C}} = \frac{\mathbb{C}}{E_0}, \,\,\,\,\,\,\, \widetilde{G}_\text{c} = \frac{G_\text{c}}{E_0L_0},  \,\,\,\,\,\,\,
    \widetilde{\underline{u}} = \frac{\underline{u}}{L_0},  \,\,\,\,\,\,\, 
    \widetilde{\ell} = \frac{\ell}{L_0},  \,\,\,\,\,\,\,
    d\widetilde{\Omega} = \frac{d\Omega}{L_0^3},  \,\,\,\,\,\,\,
    \widetilde{\doubleunderline{\epsilon}}=\doubleunderline{\epsilon},  \,\,\,\,\,\,\,
    \widetilde{\alpha}=\alpha
\end{equation}
For easier readability, the tilde above the non-dimensional parameters is dropped in what follows.

\begin{figure}[bt]
  \centering
  \includegraphics[width=0.9\linewidth]{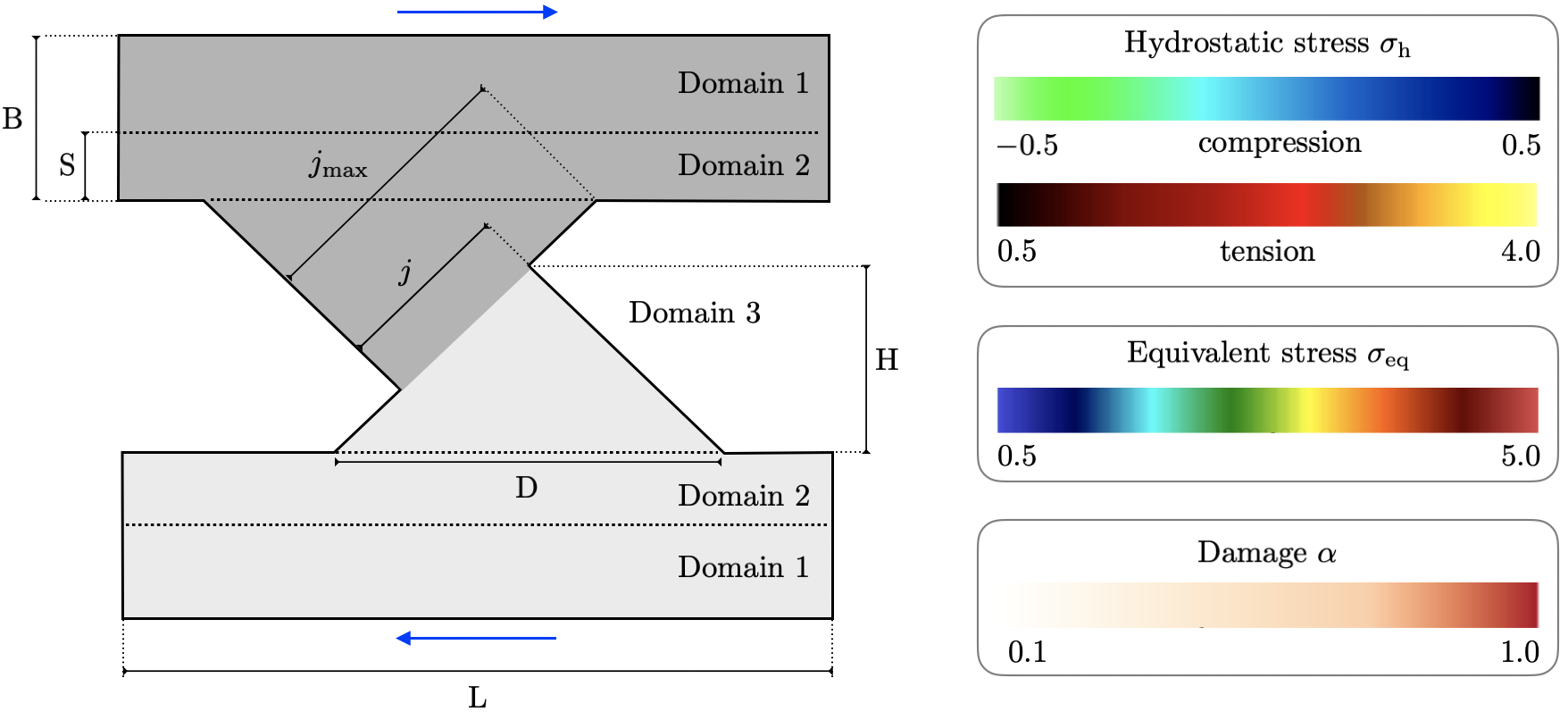}
  \caption{Schematic representation of the domain used for the simulations. Domains $1$ and $2$ form the bulk of the domain while domain $3$ represents the junction formed by the two asperities. Domains $1$ only serve to avoid boundary effects near the junction: for the sake of computational efficiency these domains has a coarser mesh size and cannot be damaged.  Domains $2$ represent the areas of the bulk where cracks are likely to propagate. They thus have the same mesh size and material parameters as the junction domain $3$. To avoid boundary effects, the bulk's height and length are set to $B\approx 3D$ and $L\approx 5D$, while the height of domains $2$ is set to $S\approx D$. Arrows represent the displacement imposed at each time step on both top and bottom horizontal boundaries. The thresholds used for the visualization of the results are displayed on the right. The stress measures are non-dimensionalized with respect to $E_0$.}
  \label{fig:mesh-length}
\end{figure}

\subsection{Implementation}
\label{sec:mesh-load-cond}
We numerically reproduce the experiment from Brockley and Fleming \cite{brockley-1965}, addressing wear and debris formation in sheared junctions.
The computational domain $\Omega\in\mathbb{R}^2$ representing the adhesive junction is comprised of two asperities in contact to each other, as shown in Figure \ref{fig:mesh-length}. Both asperities have a triangular geometry with base $D$, height $H$ and slenderness ratio $H/D$. The current and maximal values of the junction length are respectively denoted as $j$ and $j_\text{max}$. The ratio $J=j/j_\text{max}$ (with $J \in [0,1]$) is introduced to describe the portion of the asperity side which is actually involved in the formation of the junction. The adhesive strength at the junction is considered to be equal to the bulk strength and thus the domain is modeled as a continuous medium.
A quasi-static loading is imposed through the relative horizontal displacement of the top and bottom boundaries (sketched by arrows in Figure \ref{fig:mesh-length}), while the vertical displacement is set to zero. The other boundaries of the domain are free to move in both horizontal and vertical directions. The incremental value of the applied displacement is set small enough to ensure a damage-free elastic behavior at early times.

The study is performed under plane strain conditions using 2D unstructured meshes with element size $\delta$. The simulations are performed with the open source code mef90\footnote{Available at \href{https://www.bitbucket.org/bourdin/mef90-sieve}{https://www.bitbucket.org/bourdin/mef90-sieve}} \cite{bourdin-mef90}. The constraint minimization with respect to the damage $\alpha$ is done using the solvers from PETSc. The minimization with respect to the displacement field $\underline{u}$ is solved using preconditioned conjugated gradients. The numerical fracture toughness \cite{bourdin-2008} is $G_\text{c}^\text{num} = G_\text{c}\left(1+3\delta/8\ell\right)$.  According to literature \cite{ambati-2015-review,amor-2009}, a mesh size such as $\delta/\ell \in [1/4, 1/5]$ is able to represent the crack evolution without any mesh dependency while the ratio $\ell/L_0$ should tend to zero. In this study, we consider $L_0=D$ and set the non-dimensional constants
 as follows: $\delta=0.005$, $\eta=10^{-6}$, $\nu=0.2$, $E=1$ and $G_\text{c}=1$, if it is not otherwise specified. 
The base of the asperities is $D=1$. Simulations are first conducted by assuming the regularization parameter $\ell$ such that $\ell/D=1/50$ (that is, for $\ell=0.02$).

We use different color bars for the visualization of the results, as shown in Figure \ref{fig:mesh-length}.
The ranges and thresholds are set in order to provide a clear understanding of the mechanisms observed but are not meant to have any quantitative implication. The presence of a crack is highlighted by shading the region where the damage variable exceeds the threshold $\alpha\geq 0.1\%$. The equivalent stress is $\sigma_\text{eq}=\sqrt{(3/2)\doubleunderline{\sigma}_\text{d}:\doubleunderline{\sigma}_\text{d}}$, where $\doubleunderline{\sigma}_\text{d}$ is the stress deviator, whereas the hydrostatic stress is $\sigma_\text{h}=\tr \doubleunderline{\sigma}/2$. Equivalent and hydrostatic stresses are displayed by applying the thresholds $0.5\leq \sigma_\text{eq} \leq 5.0$, $-5.0 \leq \sigma_\text{h}\leq -0.5$ (in compression) and $0.5 \leq \sigma_\text{h} \leq 4.0$ (in tension).  

In order to clearly visualize crack patterns and stress fields, the figures in the sequel only display the region of the computational domain containing the junction: note that the figures show zoom-ins, cropped from a much bigger domain. Moreover, additional simulations have been conducted by homothetically increasing the size of the computational domain: identical failure mechanisms have been observed.

\begin{figure*}[htb!]
  \centering
  \begin{overpic}[width=0.45\textwidth]{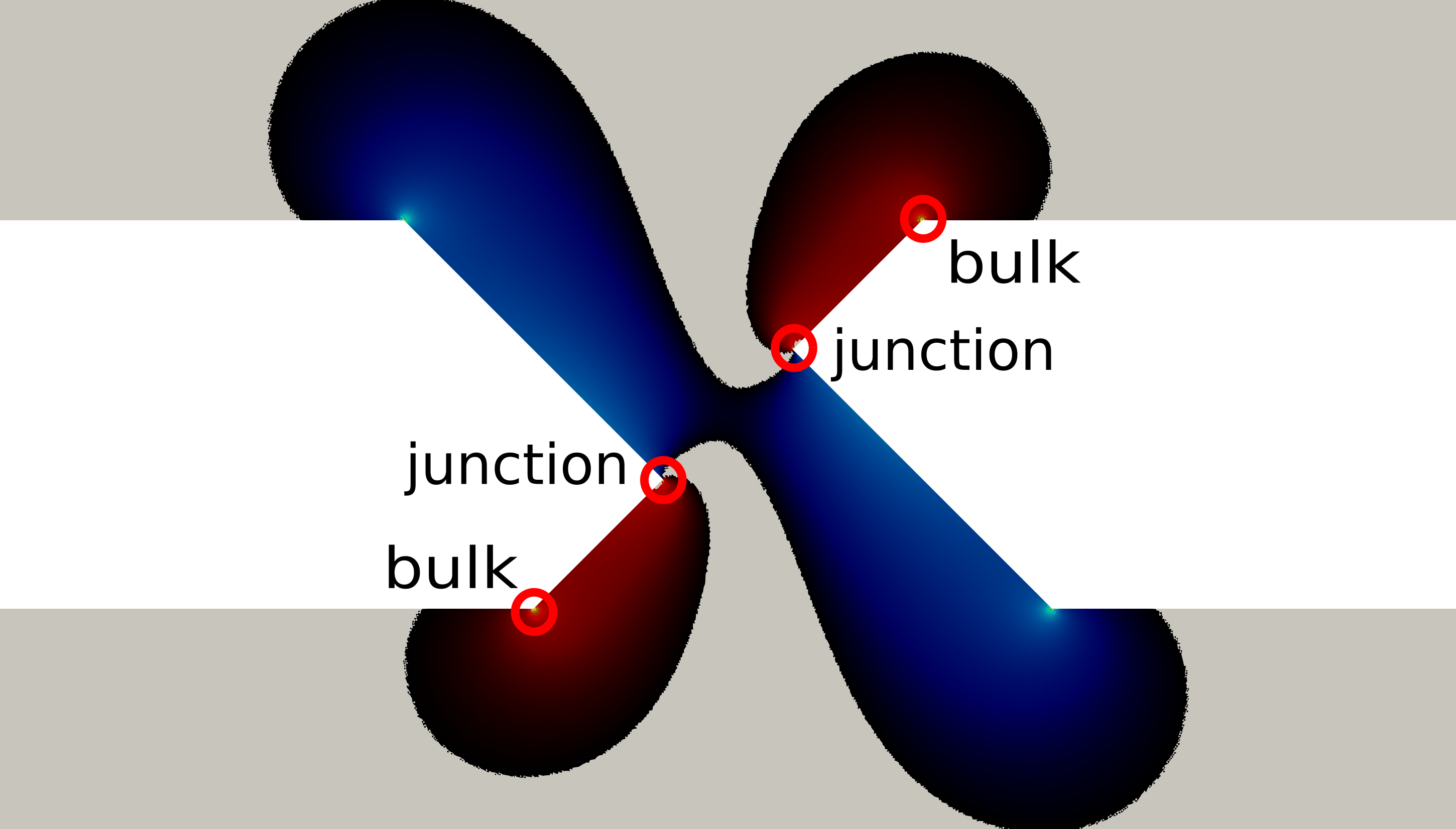}
    \put (5,50) {(a)}
  \end{overpic}\hspace{1cm}%
  \begin{overpic}[width=0.45\textwidth]{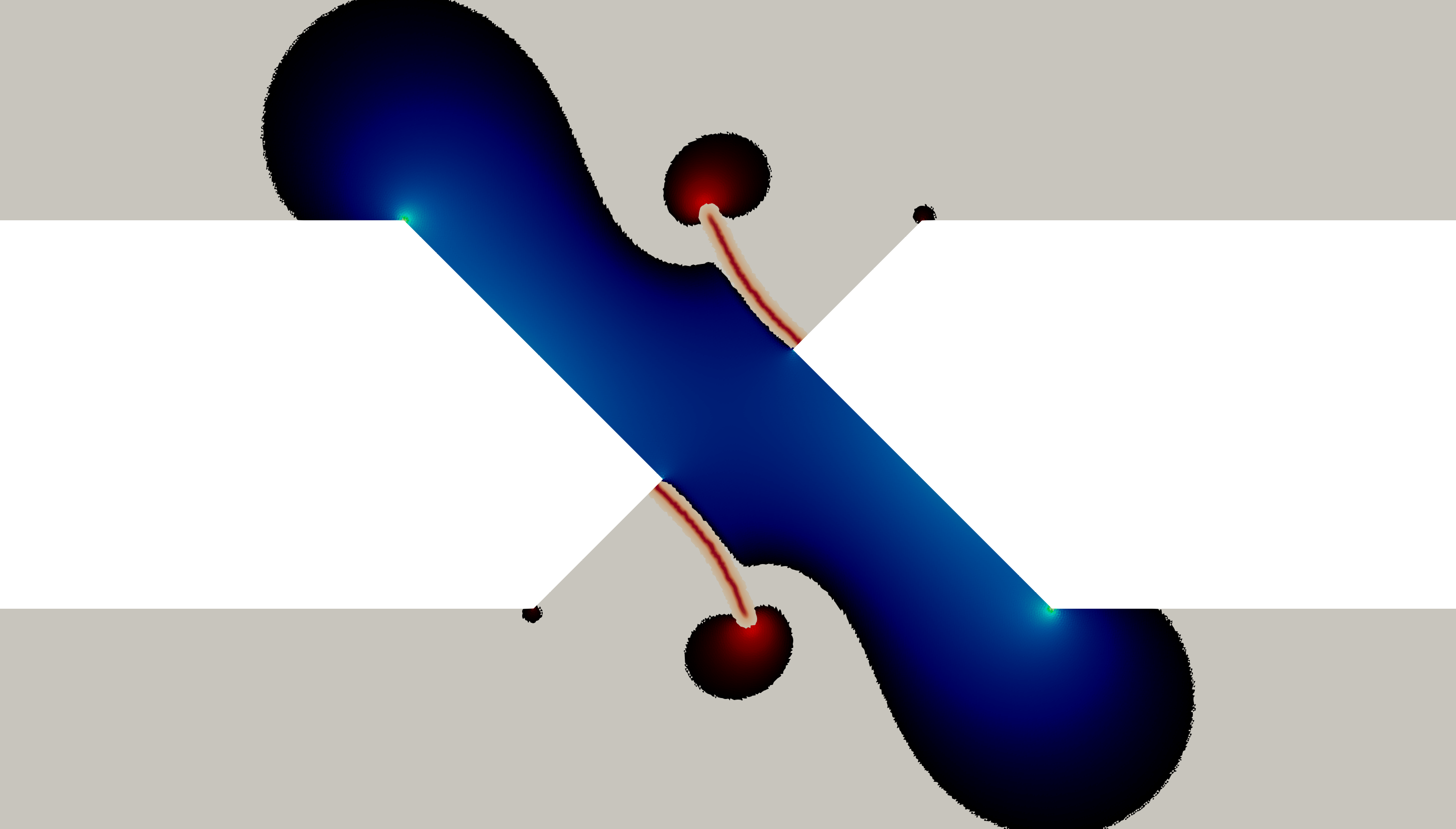}
    \put (5,50) {(b)}
  \end{overpic}
  \caption{Visualization of the crack nucleation under PH conditions for a junction with $J=0.5$ and $H/D=0.5$. The hydrostatic stresses as well as the damage field are displayed according to the thresholds in Figure \ref{fig:mesh-length}. (a) Time-step $=58$: the undamaged state shows both compressive and tensile stress concentrations. The red circles indicate where tensile stresses concentrate, which correspond to potential crack nucleation spots. (b) Time-step $=59$: the damaged state computed at the next time-step shows a pair of cracks nucleated inside the junction (at the 'junction' stress concentration). The nucleation releases the tensile stresses while compressive ones remain and deflect the crack path. Note that the figures show zoom-ins, cropped from a much bigger domain.}
  \label{fig:posH-nuc}
\end{figure*}

\begin{figure*}[ht!]
  \centering
  \begin{overpic}[width=0.45\textwidth]{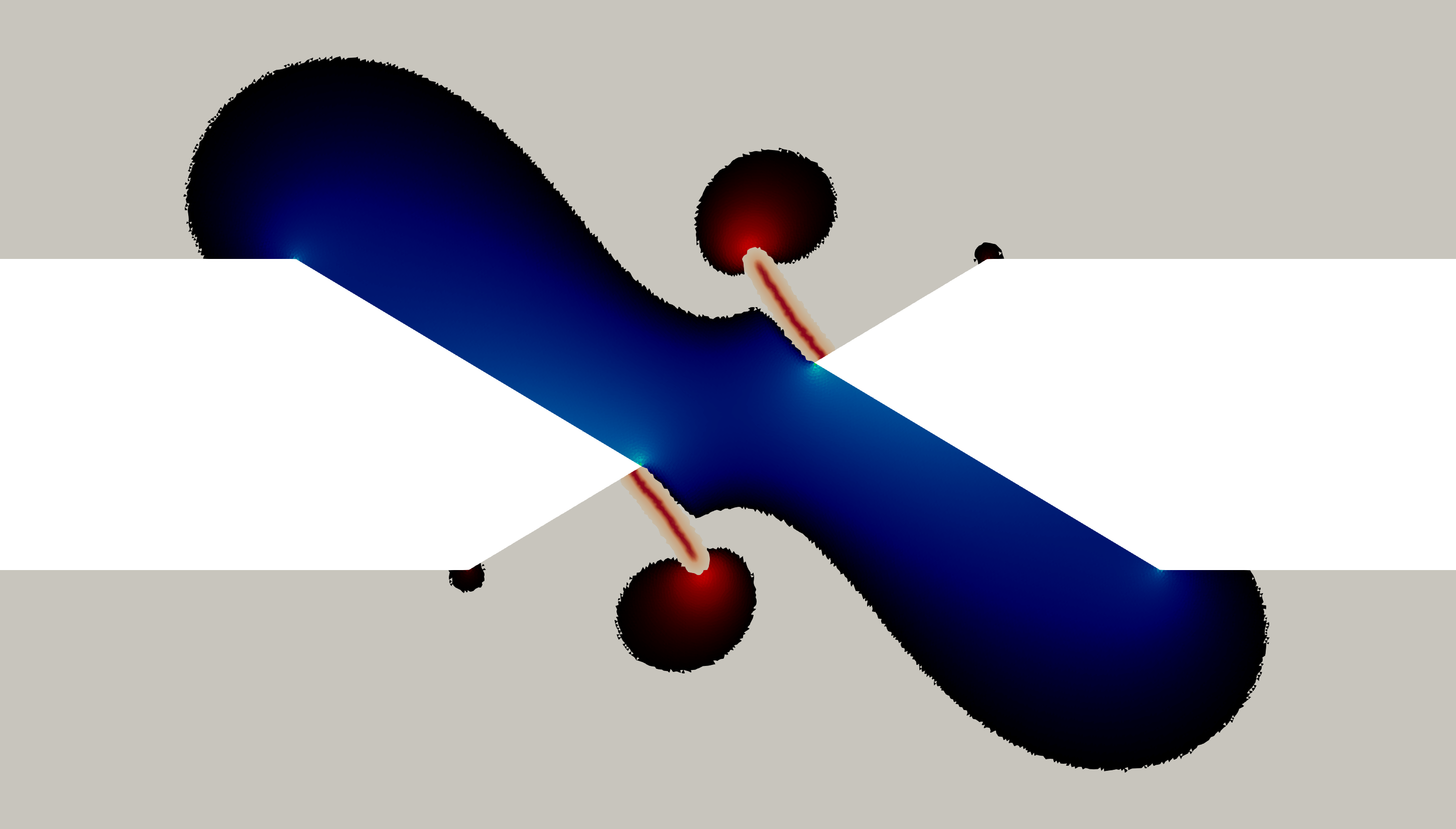}
    \put (5,50) {(a)}
  \end{overpic} \hspace{1cm}%
  \begin{overpic}[width=0.45\textwidth]{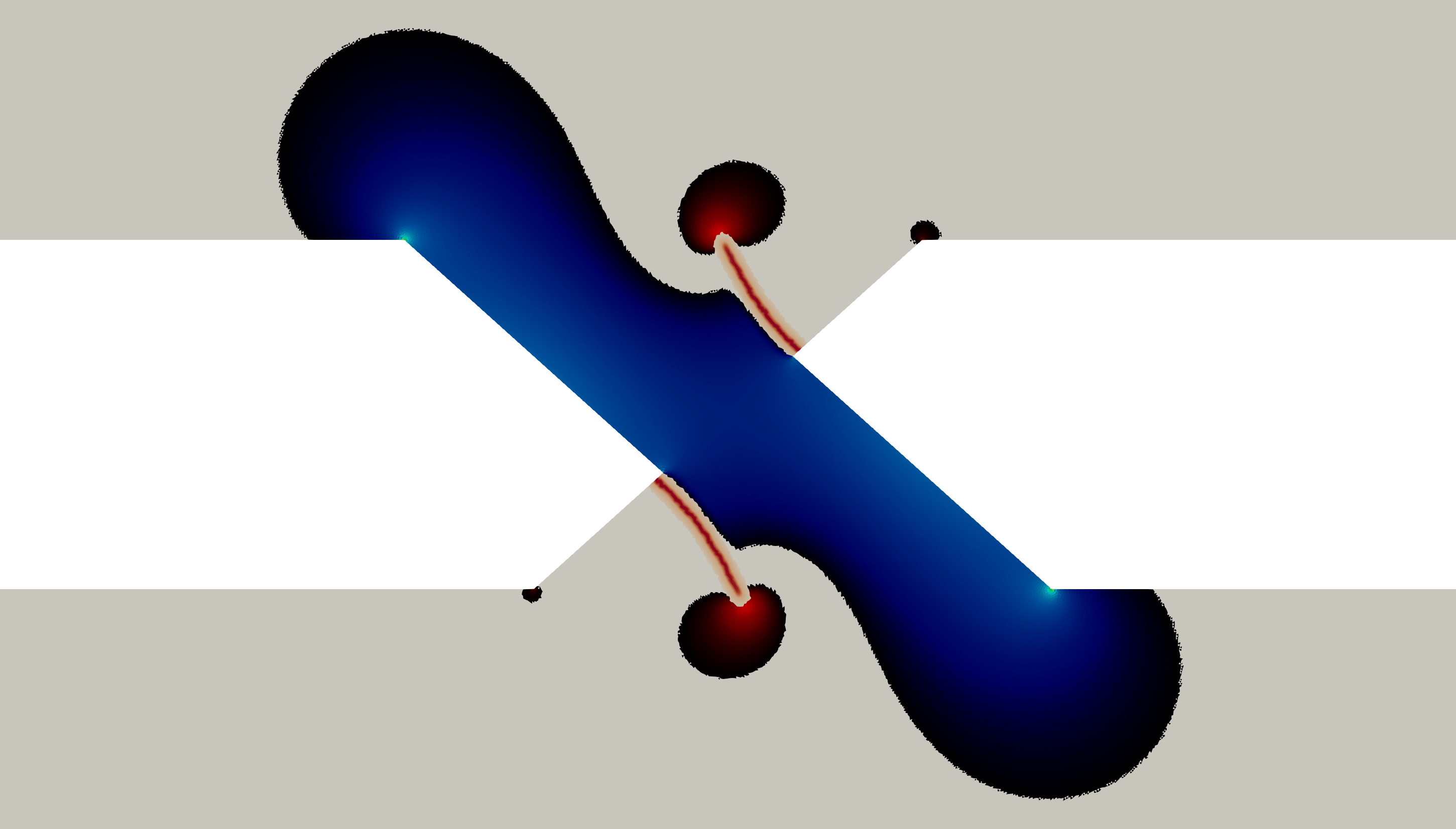}
    \put (5,50) {(b)}
  \end{overpic} \\
  \vspace{1cm}
  \begin{overpic}[width=0.45\textwidth]{el_pos-hydro_ah05_J05_hydro_59.png}
    \put (5,3) {(c)}
  \end{overpic} \hspace{1cm}%
  \begin{overpic}[width=0.45\textwidth]{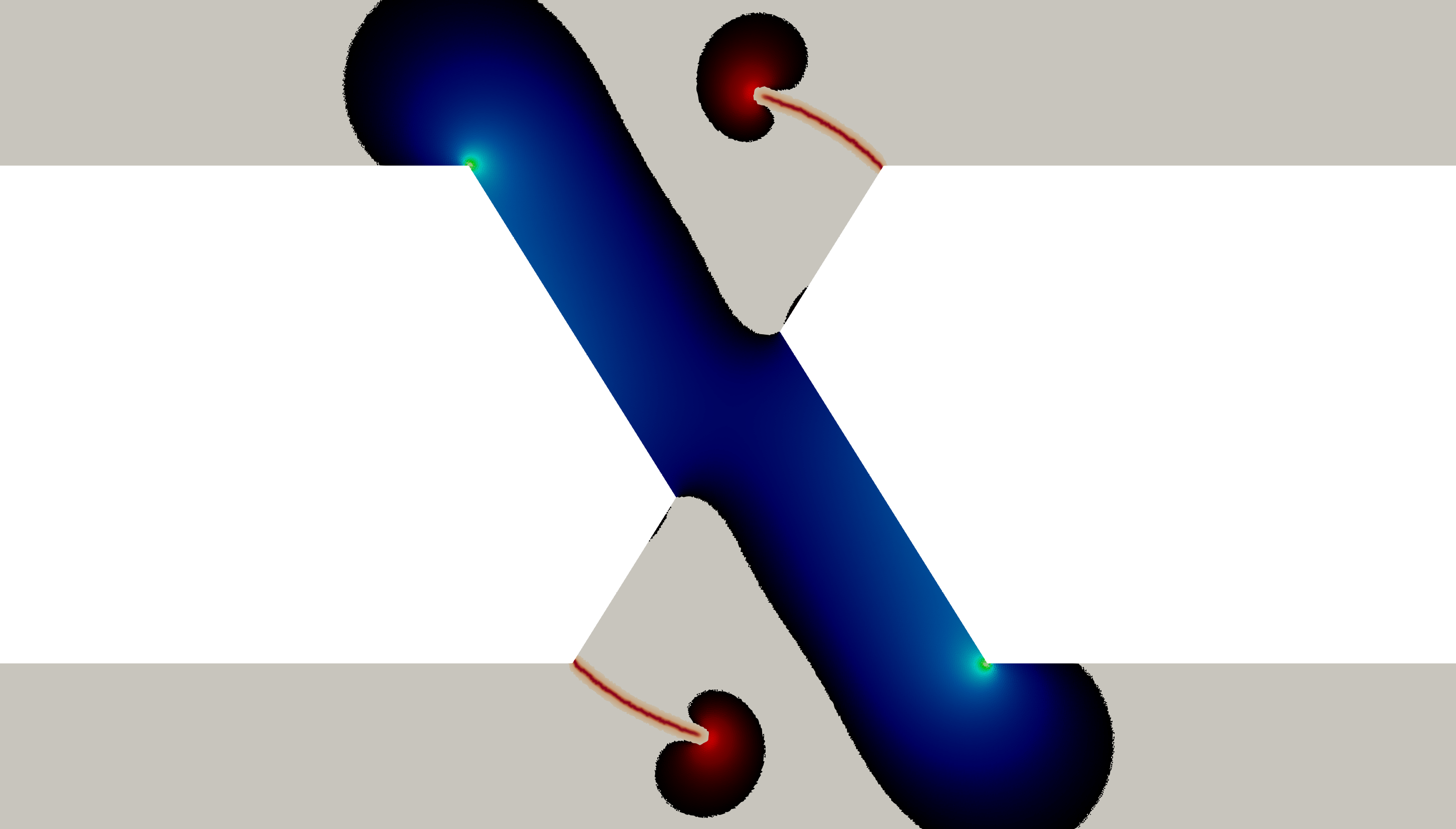}
    \put (5,3) {(d)}
  \end{overpic}
  \caption{Visualization of the transition in the crack nucleation pattern in single junctions under PH conditions, as a consequence of a change in the ratio $H/D$ with constant junction parameter $J=0.5$. The hydrostatic stresses and damage are displayed according to the thresholds in Figure \ref{fig:mesh-length}. (a) $H/D= 0.3$. (b) $H/D=0.45$ and (c) $H/D = 0.5$: a pair of cracks nucleate inside the junction. The cracks path is deflected towards the bulk by the high compressive stresses developed in the junction. (d) $H/D=0.8$: a pair of cracks nucleates in the bulk and grows along the asperity basis. Note that the figures show zoom-ins, cropped from a much bigger domain.}
  \label{fig:posH-mech}
\end{figure*}

\section{Fracture analysis of single-asperity junctions}
\label{sec:single-junctions}

\subsection{Positive-hydrostatic model}
\label{sec:posH}
We first perform a set of simulations using the positive-hydrostatic (PH) constraint in Eq.\,\eqref{eq:posH}, which implies that a crack can nucleate only as a consequence of tensile stresses. 

Figure \ref{fig:posH-nuc} (a) shows the distribution of compressive and tensile hydrostatic stresses at the time-step which precedes the nucleation of cracks. We observe four locations with tensile stress concentrations, which we collect in two groups. The first one, labeled as `junction', corresponds to the corners formed by the incomplete overlap of the asperities. The second one, labeled as `bulk', corresponds to the corners where the asperity rises from the bulk. Indeed, for $J=1$, the two stress concentrations coincide as the overlap of the asperities is complete. Results also reveal the occurrence of compressive hydrostatic stresses, which develop from the junction -where the two asperities are pressed against each other- and concentrate at the opposite side of the `bulk' corners. 
At the time-step immediately after (Figure \ref{fig:posH-nuc} (b)), two cracks nucleate at the `junction' stress concentrations and the tensile hydrostatic stresses are released through fracture.

The influence of the geometry of the asperities on crack nucleation is studied in Figure \ref{fig:posH-mech} by varying the slenderness $H/D$, for a given value of the junction parameter $J$. Results clearly show a shift in the crack nucleation spot from the `junction' to the `bulk' as the asperities become more slender, thus suggesting a larger loss of material at complete failure. We believe that this can be directly related to the resulting wear particle size, although the latter cannot be precisely quantified as the final state of the adhesive junction is not accessible by the model at hand. 
The reason for this is twofold. 
First, our approach relies on small strains, whereas the failure of a sheared adhesive junction is  believed to be also influenced by the large displacements that take place \cite{milanese-2019}. Second, it can be clearly seen in Figures \ref{fig:posH-nuc} and \ref{fig:posH-mech} that the compressive stress field deflects the crack paths towards the bulk. We believe that this is an artifact of the PH model, as in material regions where the volume change is negative the hydrostatic strain energy cannot be released through fracture. The cracks then deviate towards the bulk, instead of propagating throughout the asperities and eventually detaching a debris particle. 

 Nonetheless, MD simulations  \cite{aghababaei-2016, aghababaei-2017, aghababaei-2018, brink-2019} showed that the location of the damage at the early stages of the shearing process provides a good indication on the final state of the junction at failure. In other words, one can assume that a crack nucleating in the bulk will very likely detach a debris particle, whereas  failure mechanisms such as slip will most probably not.
As such, we associate the formation of two cracks at the `junction' as a mechanism that would eventually lead to the formation of a \textit{small particle}, as the area around the adhesive interface gets separated from the rest of the domain by the cracks (e.g., as in  Figure \ref{fig:posH-mech} (a), (b) and (c)). This under the assumption that without the cracks being deflected by the compressive stress field, they would rather propagate in the junction than towards the bottom part of the domain. On the other hand, when a crack nucleates at the `bulk', its path suggests the detachment of the whole junction as a single debris: Figure \ref{fig:posH-mech} (d) shows how the cracks nucleate at the `bulk' stress concentrations and subsequently propagate at the asperity base,  turning as they are about to detach the full junction.  We  thus associate this kind of crack nucleation to the formation of a \textit{large particle}.

\begin{figure}[hb!]
  \centering
  \includegraphics[width=0.7\linewidth]{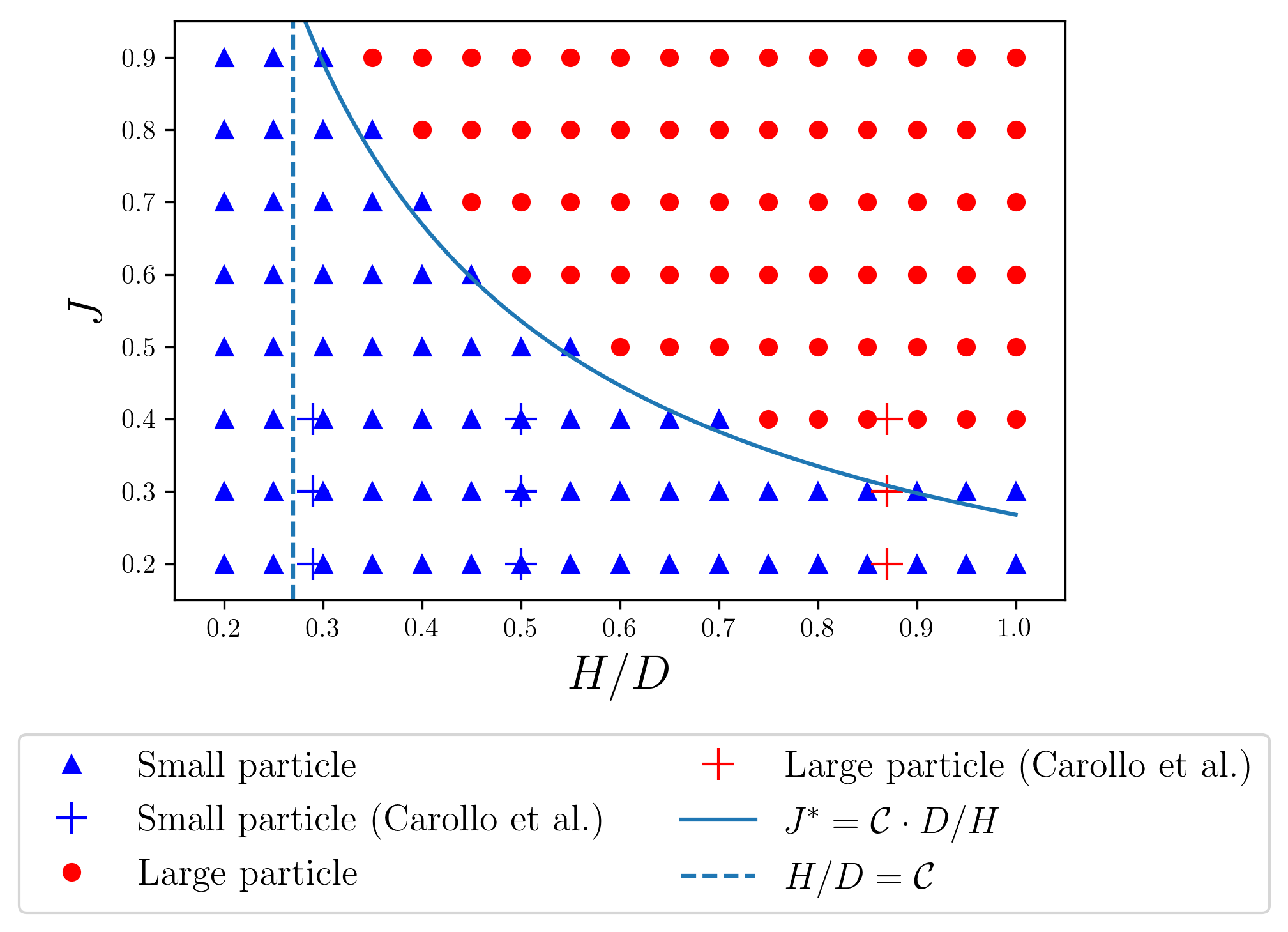}
  \caption{Evolution of the failure mechanism under PH conditions, as a function of the domain's geometry where each point represents a different simulation. The blue triangles represent junctions that failed through cracks nucleation inside the junction as shown in Figure \ref{fig:posH-mech} (a), (b) and (c). Red dots represent junctions that failed through the nucleation of a crack in the bulk as shown in Figure \ref{fig:posH-mech} (d). The solid line is a fitting on the points marking the transition from one regime to the other. The crosses represent data points from a similar study \cite{carollo-2019} which uses a slightly different phase-field formulation. $\mathcal{C} = 0.27$.}
  \label{fig:posH-chart}
\end{figure}

The complete study of the geometry's effect on the failure of the adhesive junction is reported in Figure \ref{fig:posH-chart}, in terms of the slenderness $H/D$ and the junction parameter $J$. We observe that the transition from small to large particle detachment follows a clear trend described by a law of the form
\begin{equation}
  \label{eq:J-star}
  J^* = \frac{D}{H}\, \mathcal{C}
\end{equation}
The latter intersects the dashed line $H/D = \mathcal{C}$ when $J\rightarrow1$, which corresponds to the limit case of a junction formed by completely overlapping asperities.
For the material parameters used in this study, the fitting coefficient in Eq.\,\eqref{eq:J-star} is $\mathcal{C} \approx 0.27$.
Additional simulations performed on domains where all lengths are doubled while keeping the same $\delta$ and $\ell$ further confirmed that the transition from small to large particle formation is governed by the slenderness $H/D$ rather than by the height of the asperity only.

The results from Carollo \textit{et al.} \cite{carollo-2019} are also reported in Figure \ref{fig:posH-chart}, blue and red crosses respectively denoting small and large particle formation. Authors \cite{carollo-2019} used a phase-field formulation which differs from ours for the splitting of the elastic strain energy, as well as slightly different boundary conditions. The number of available simulations is not large enough for us to properly compare the outcomes of the two studies, however we do notice a certain agreement in the prediction of small particle formation whereas results tend to disagree when large particles are detached.

\begin{figure*}[!t]
  \centering   
  \begin{overpic}[width=0.45\textwidth]{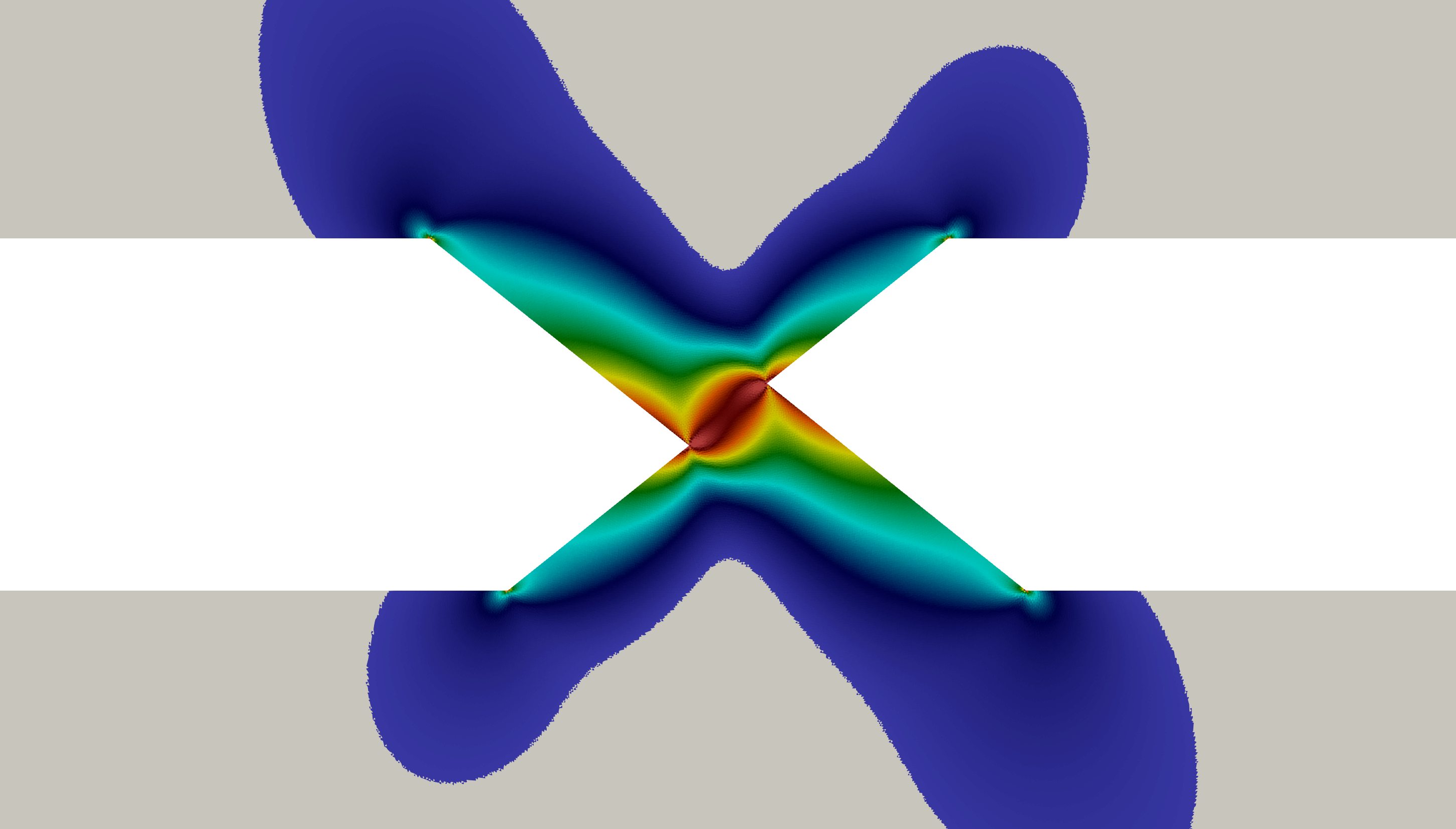}
    \put (5,50) {(a)}
  \end{overpic}\hspace{1cm}%
  \begin{overpic}[width=0.45\textwidth]{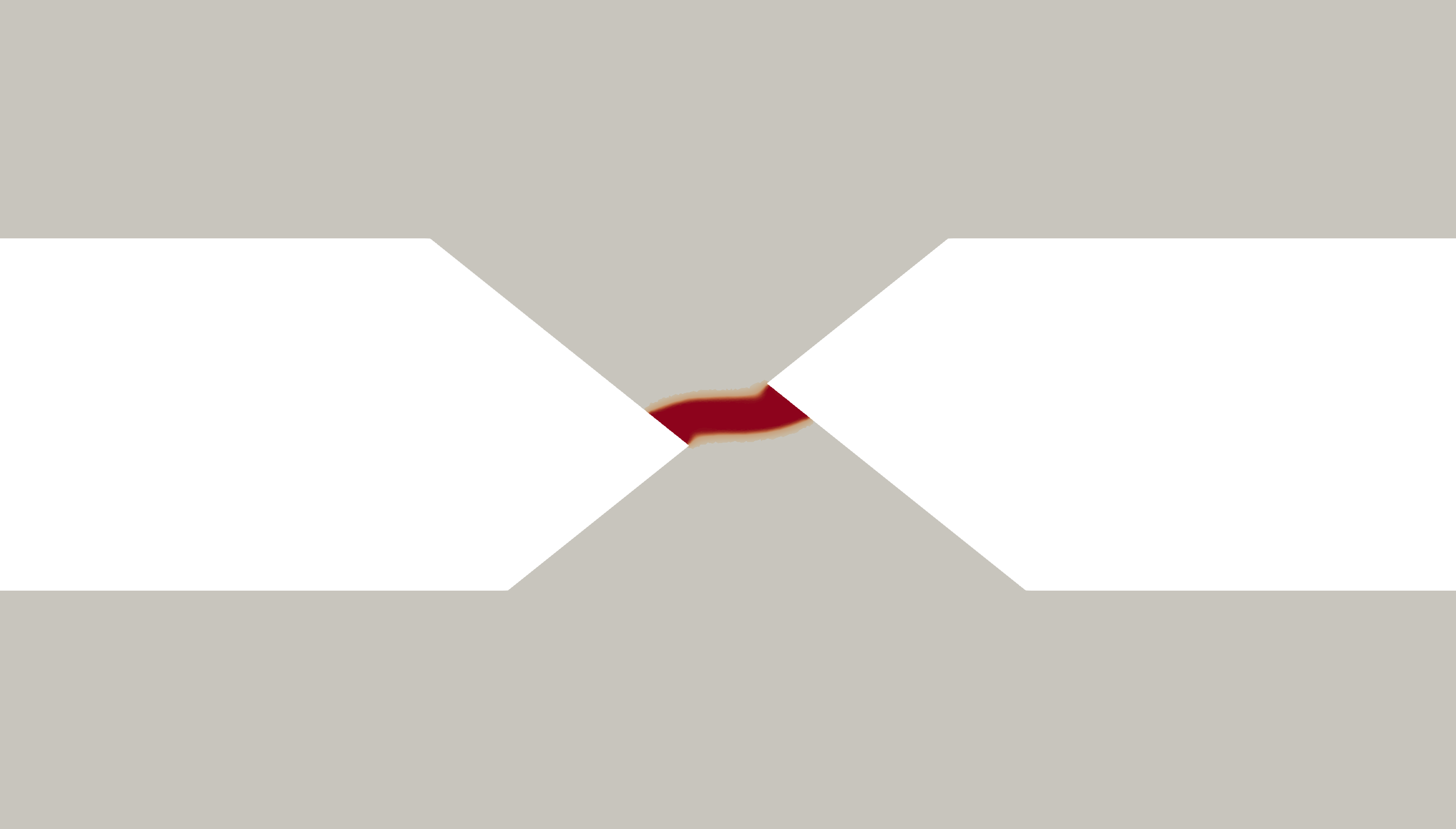}
    \put (5,50) {(b)}
  \end{overpic}
  \caption{Visualization of the junction failure through a slip mechanism under HD conditions, for a junction with $J=0.3$ and $H/D=0.4$. The equivalent stress as well as the damage field are displayed according to the thresholds in Figure \ref{fig:mesh-length}.  (a) Time-step$=38$: the domain is still intact, the highest values of equivalent stress are located along the interface between the two asperities.  (b) Time-step$=39$: a shear band develops along the interface and the stresses are completely released. Note that the figures show zoom-ins, cropped from a much bigger domain.}
  \label{fig:hydroD-nuc}
\end{figure*}

\begin{figure*}[!h]
  \centering
  \begin{overpic}[width=0.45\textwidth]{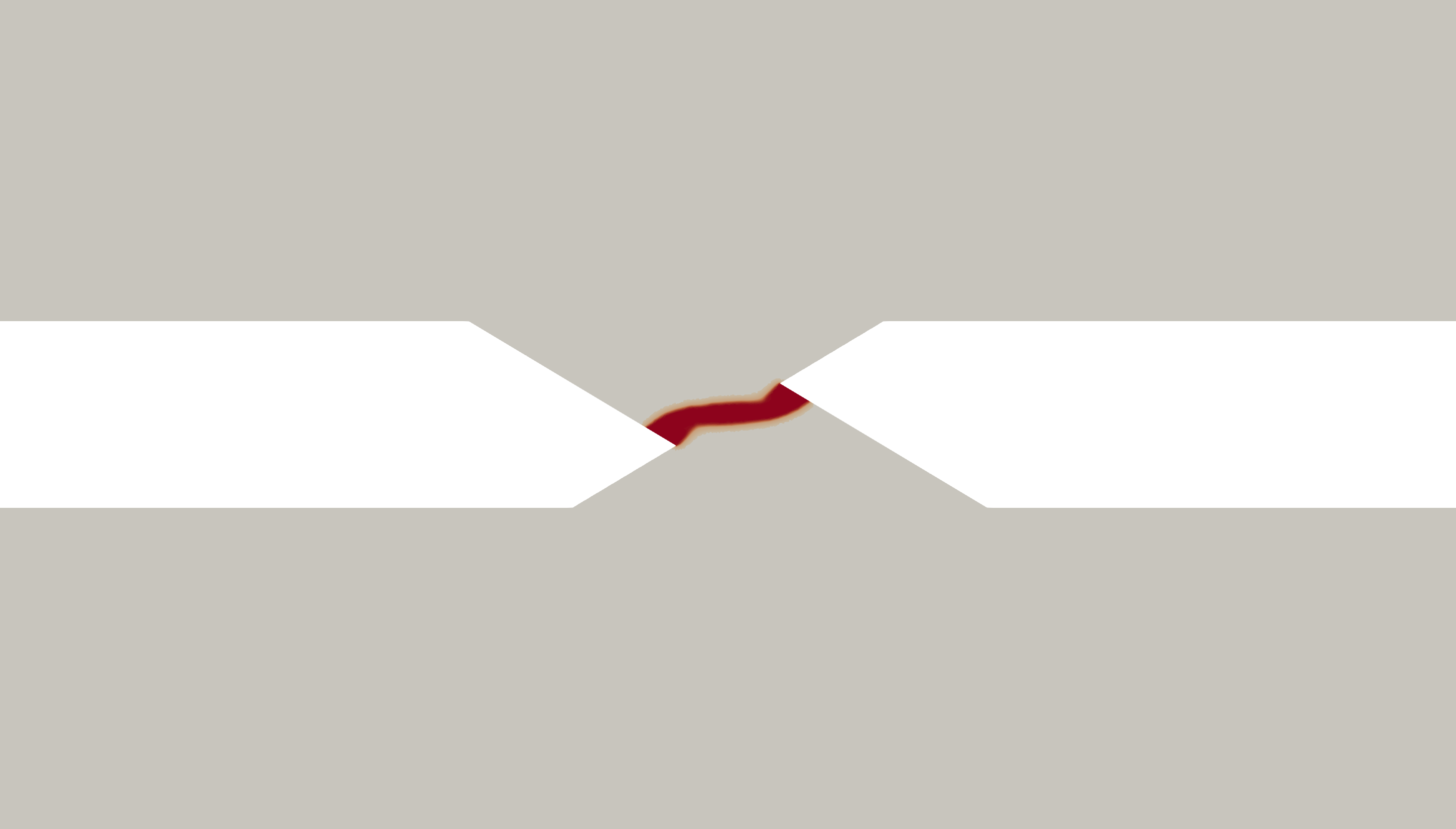}
    \put (5,50) {(a)}
  \end{overpic} \hspace{1cm}%
   \begin{overpic}[width=0.45\textwidth]{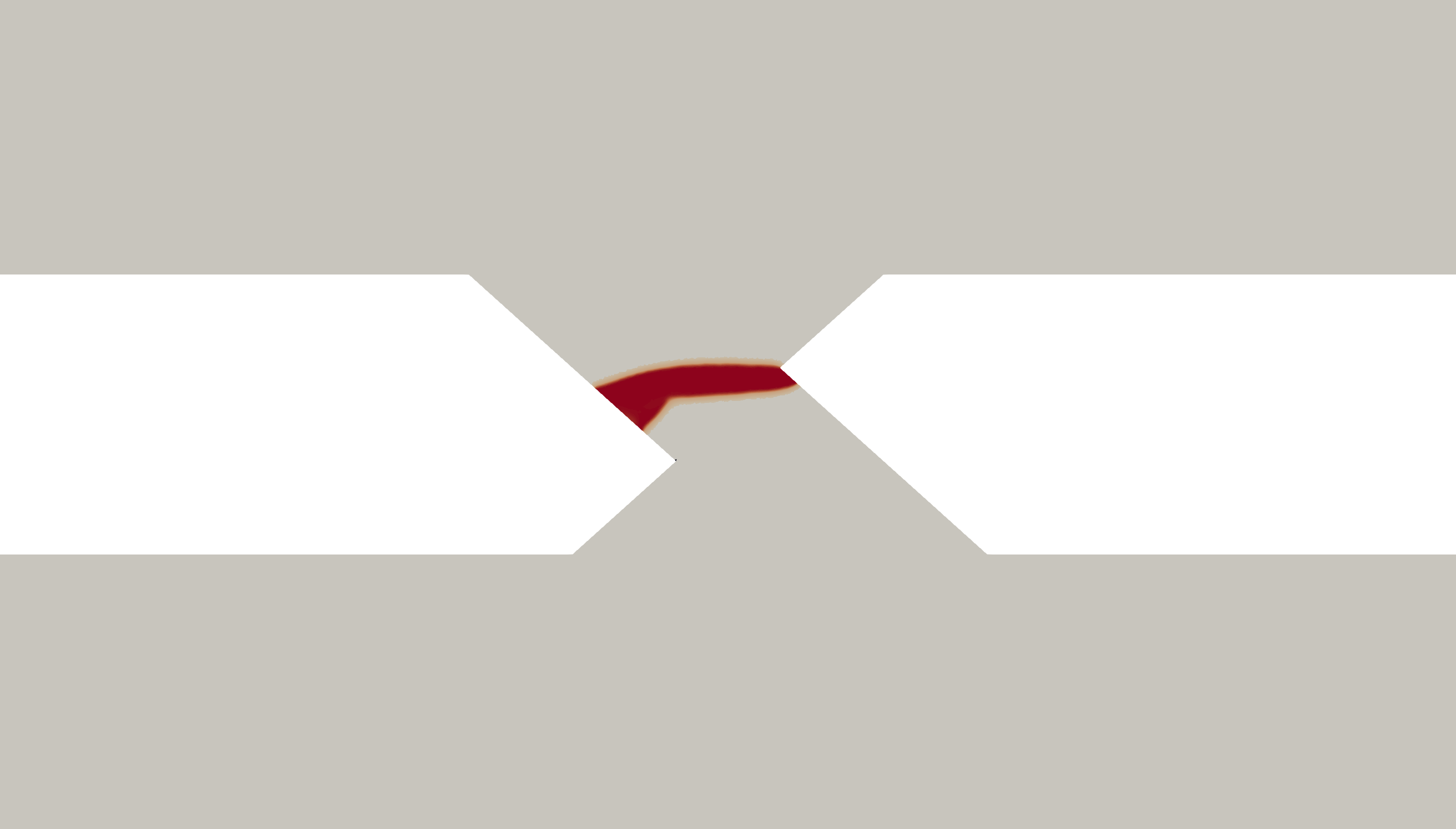}
    \put (5,50) {(b)}
  \end{overpic} \\
  \vspace{1cm}
  \begin{overpic}[width=0.45\textwidth]{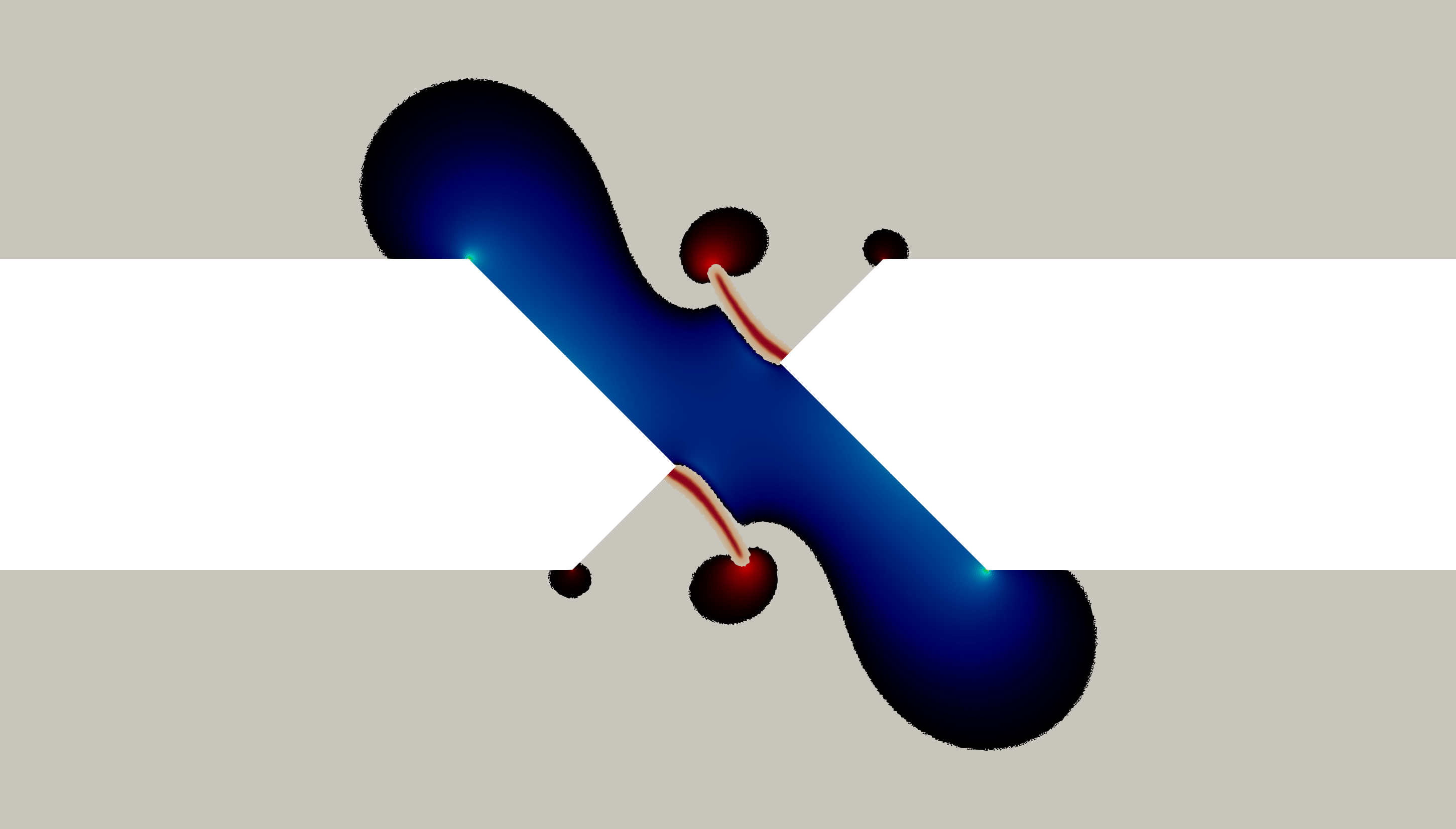}
    \put (5,4) {(c)}
  \end{overpic} \hspace{1cm}%
  \begin{overpic}[width=0.45\textwidth]{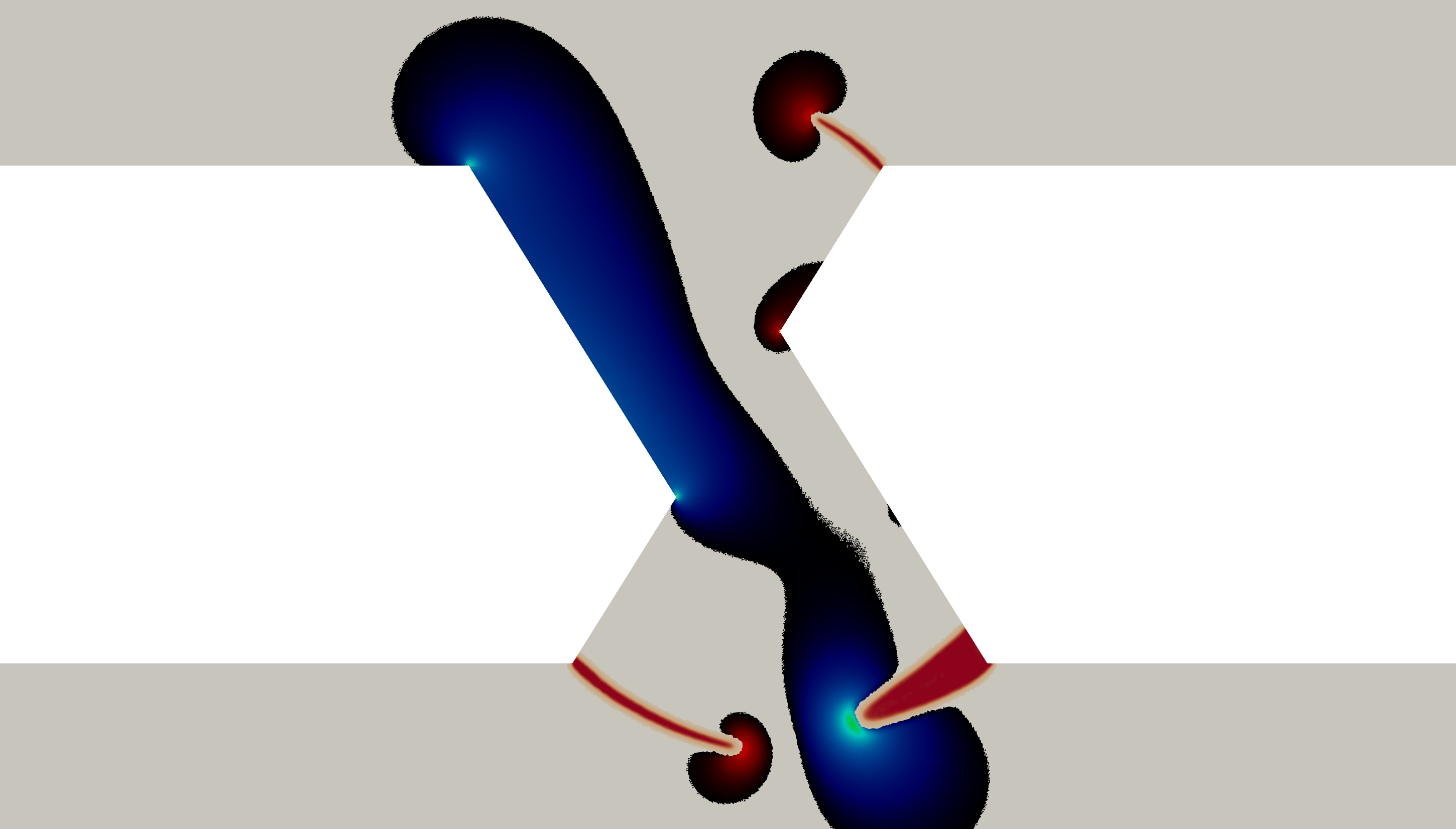}
    \put (5,4) {(d)}
  \end{overpic}
  \caption{Visualization of the transition in the crack nucleation pattern in single junctions under HD conditions, as a consequence of a change in the ratio $H/D$ with constant junction parameter $J=0.5$. The hydrostatic stresses and damage fields are displayed according to the thresholds in Figure \ref{fig:mesh-length}. The geometries are the exact same as in Figure \ref{fig:posH-mech} (PH model). (a) $H/D= 0.3$: a single shear band nucleates along the interface between the two asperities. (b) $H/D = 0.45$: a single shear band nucleates horizontally from one of the stress concentration in the junction. (c) $H/D=0.5$:  a pair of cracks nucleates inside the junction. The cracks path is deflected towards the bulk by the high compressive stresses developed in the junction.  (d) $H/D=0.8$: a pair of cracks nucleates in the bulk along the asperities' basis. A shear band also develops from the opposite corner, suggesting a large particle detachment. Note that the figures show zoom-ins, cropped from a much bigger domain.}
  \label{fig:hydroD-mech}
\end{figure*}

\begin{figure}[hbt!]
  \centering
  \includegraphics[width=0.65\linewidth]{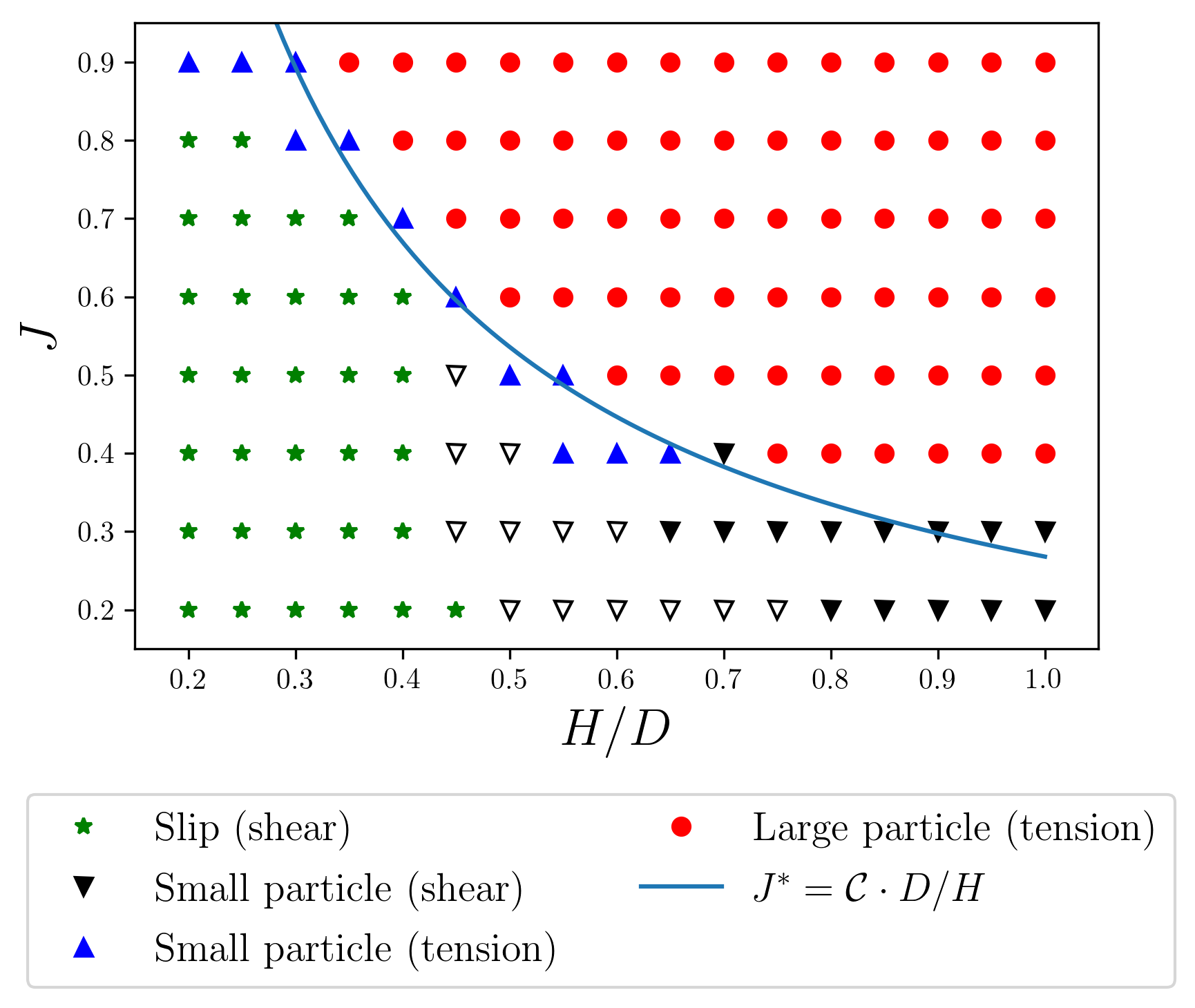}
  \caption{Evolution of the failure mechanisms under HD conditions, as a function of the domain's geometry, where each point represents a different simulation. The green stars represent the junctions that failed through a slip mechanism, as shown in Figure \ref{fig:hydroD-mech} (a). The black downward triangles represent the junctions that failed by developing one (hollow markers) or two (dense markers) straight shear bands, the first case being shown in Figure \ref{fig:hydroD-mech} (b). The blue upward triangles represent junctions that failed through crack nucleation inside the junction, as shown in Figure \ref{fig:hydroD-mech} (c). The red dots represent junctions that failed through the nucleation of cracks in the bulk, as shown in Figure \ref{fig:hydroD-mech} (d).  The solid line is the exact same function as show in Figure \ref{fig:posH-chart} and it now marks the transition from small or no-particle formation to large particle detachment. $\mathcal{C} = 0.27$.}
  \label{fig:hydroD-chart}
\end{figure}

\subsection{Hydrostatic-deviatoric model}
\label{sec:hydroD}
We now consider the hydrostatic-deviatoric (HD) model in Eq.\,\eqref{eq:hydroD}, which allows fracture to occur as a consequence of both shear and positive hydrostatic stresses. This change in the coupling between strain and fracture energy brings failure patterns that  cannot be captured using a PH model. In fact, in addition to tension-driven cracks as described in the previous Section, some geometries exhibit failures that involve shear bands (see Figure \ref{fig:hydroD-nuc}). These regions are characterized by the presence of a diffuse damage and thus are thicker than tensile cracks \cite{amor-2009}.

With respect to what observed under PH conditions, our simulations highlight the occurrence of two additional failure modes for the adhesive junction: \textit{slip} and \textit{shear}. Slip describes a shear band that forms along the interface between the two asperities, while shear stands for shear bands that develop horizontally.
The major difference is that the second mechanism leads in certain cases to the detachment of a small particle (in the same fashion as described in Section \ref{sec:posH}), whereas the first does not. 

The transition across the four types of failure as the slenderness $H/D$ of the asperities increases is shown in Figure \ref{fig:hydroD-mech}, for a given value of the junction parameter $J$. The geometries are the same as those analyzed in Figure \ref{fig:posH-mech}. For small values of $H/D$, a shear band forms at the junction interface and the asperities slip past one another (Figure \ref{fig:hydroD-mech} (a)). An increase in $H/D$ results in a shear band running throughout the asperity parallel to the macroscopic shear direction (Figure \ref{fig:hydroD-mech} (b)). 
As the junction becomes even more slender, cracks nucleate hinting to small (Figure \ref{fig:hydroD-mech} (c)) and large (Figure \ref{fig:hydroD-mech} (d)) particle detachment.

The complete study of the failure modes of the adhesive junction is reported in Figure \ref{fig:hydroD-chart}, as a function of the slenderness $H/D$ and the junction parameter $J$. Interestingly, the exact same law in Eq.\,\eqref{eq:J-star} postulated for describing the transition between small and large particle detachment  still holds under HD constraints.  
The difference from Figure \ref{fig:posH-chart} to Figure \ref{fig:hydroD-chart} thus lies in the area of the chart where there is either small or no particle detachment.  
In fact the left side of the chart, where junctions are relatively small and asperities bulky, is now occupied by the slip mechanism. We believe that this failure mode is a consequence of both the mild slope of the interface between the asperities and the relative proximity of the two stress concentrations inside the junction. On the other hand, as the asperities' slenderness $H/D$ increases, the slope of the interface rises and the shear bands grow horizontally through the junction rather than along the interface.  For the most slender asperities, we even observe two parallel shear bands that detach a small particle from the junction.

More generally, we observe that almost all simulations that lie under the transition curve are now showing failure mechanism which involve shear bands rather than cracks triggered by opening stresses. The only simulations that still present tension-driven cracks inside the junction (blue upside triangles) are located in the proximity of the transition curve.
Hence, our results indicate that the transition from small or no-particle (slip) to large-particle formation is associated with a change from shear bands to tension-driven cracks.

\subsection{On the existence of a critical length scale}\label{sec:theory}
The transition from shearing an asperity with negligible mass loss to the detachment of a large debris particle has already been observed in MD simulations \cite{aghababaei-2016}. The first mechanism led to the ductile failure of the junction, via the progressive smoothing of the asperities involved in the sliding contact. On the other hand, brittle failure occurred in asperities with larger junction size, eventually producing a wear debris. Authors explained the transition between these two regimes by the introduction of a critical length scale $d^\ast$ (see Eq.\,(3) in \cite{aghababaei-2016}), below (resp., above) which ductile (resp., brittle) failure occurred.

Our phase-field simulations further demonstrate the existence of such a critical length-scale.
In fact, we find that the transition from relatively low mass loss to large particle formation can be captured by the law in Eq.\,\eqref{eq:J-star}. For a given slenderness $H/D$ of the asperity and material properties, Eq.\,\eqref{eq:J-star} uniquely identifies a critical junction length $J^\ast$ above which the junction detaches as a large wear debris. On the other hand, junction's lengths smaller than the critical value $J^\ast$ lead to small or no particle formation. Importantly, the transition curve \eqref{eq:J-star} is \textit{exactly the same} for both PH and HD models, as the value obtained for the constant parameter $\mathcal{C}$ is the same in both formulations.

The transition law \eqref{eq:J-star} can be analytically derived based on the arguments presented in \cite{aghababaei-2016}. 

We start by noticing that the regularization length $\ell$ of the phase-field model defines the nucleation stress
\begin{equation}
\label{eq:sigC}
    \sigma_\text{c}^2=\frac{3G_\text{c}E}{8\ell(1-\nu^2)}
\end{equation}
which corresponds to the tensile strength of the material under uniaxial loading and in plane strain conditions \cite{amor-2009}. 
Comparisons with experiments \cite{tanne-2018} demonstrated that, by choosing the regularization length $\ell$ such that $\sigma_\text{c}$ matches the tensile strength, the phase-field formulation is able to capture crack nucleation in a variety of materials and geometries.
Moreover, by recasting the variational HD formulation in a thermodynamic framework, crack nucleation can be shown to be governed by the stress criterion
\begin{equation}
\label{eq:F}
    \mathcal{F}\left(\bm{\sigma}\right)=\sigma_\text{eq}^2 + \frac{3\mu}{K}\sigma_\text{m+}^2-\frac{3\mu}{E}\sigma_\text{c}^2
\end{equation}
where $\sigma_\text{eq}$ (resp., $\sigma_\text{m+}$) is the equivalent deviatoric (resp., positive hydrostatic) stress invariant and $\sigma_\text{c}$ is the tensile strength in Eq.\,\eqref{eq:sigC}. The PH formulation leads to a criterion similar to that in Eq.\,\eqref{eq:F},  expressed only in terms of the positive hydrostatic stress $\sigma_\text{m+}$.

Our numerical computations showed that, for a given $\ell$ and depending on the geometry of the junction, the formation of a wear debris can be  triggered by tensile stresses $\sigma_\text{m+}$ or shear stresses $\sigma_\text{eq}$ (see Figure \ref{fig:hydroD-chart}). Based on the fracture criterion \eqref{eq:F}, in both cases the elastic energy $\mathcal{E}_\text{el}$ released by the detachment of a wear debris can be approximately computed as\footnote{Note that here we make the assumption that prior to failure the stress distribution in the junction is relatively uniform and that failure is triggered by either one of the two stress invariants. In this sense, the resulting expression for $\mathcal{E}_\text{el}$ is an estimate of the elastic energy released upon the detachment of a wear debris.}
\begin{equation}
\label{eq:Eel}
    \mathcal{E}_\text{el}=2 A_\text{deb} \frac{\sigma_{c}^2}{E}
\end{equation}
where $A_\text{deb}$ is the surface area of each of the particles forming the wear debris.
On the other hand, the fracture energy related to the formation of two cracks of length $L_\text{crack}$ reads as
\begin{equation}
    \mathcal{E}_\text{fr}=2 L_\text{crack} G_\text{c}
\end{equation}

Assume that an asperity junction with given length $J^\ast$ and slenderness $H/D$ fails by the detachment of two triangular debris, as a consequence of the nucleation of two horizontal cracks at the `junction' stress concentrations in Figure \ref{fig:posH-nuc} (a). The surface area of each debris particle is then $A_\text{deb}=\alpha (J^\ast)^2 H D/2$, while the crack length is $L_{crack}=\beta J^\ast D$. Coefficients $\alpha$ and $\beta$ account for the fact that the actual shape of the detached debris might be other than triangular, and that the crack path might be not parallel to the asperity's basis. 
Following \cite{aghababaei-2016} we note that, in order for a wear debris to be detached, the elastic energy stored in the junction  must  overcome the fracture energy, that is $\mathcal{E}_\text{el}\geq \mathcal{E}_\text{fr}$. This energy condition yields
\begin{equation}
\label{eq:theory_sigc}
    J^\ast\geq \frac{2\beta}{\alpha}\,\frac{E G_\text{c}}{\sigma_\text{c}^2}\,\frac{1}{H}
\end{equation}
which, in view of Eq.\,\eqref{eq:sigC}, results in
\begin{equation}
\label{eq:theory_ell}
    J^\ast\geq \frac{16\beta}{3\alpha}
    \left(1-\nu^2 \right)\,\frac{\ell}{H}
\end{equation}
Finally, by recalling that the regularization length $\ell$ has been defined as a \textit{small fraction} of the asperity's size $D$ (i.e., $\ell=D/50$ in the performed calculations), Eq.\,\eqref{eq:theory_ell} can be recast as
\begin{equation}
\label{eq:theory}
    J^\ast\geq \frac{D}{H}\,\,\mathcal{C}
\end{equation}
where constant $\mathcal{C}$ collects all the non-geometrical parameters and accounts for some of the introduced assumptions (shape of debris particle and crack paths, uniform stored energy). 

Hence, for an asperity junction with given slenderness $H/D$ and ratio $\ell/D$, Eq.\,\eqref{eq:theory} identifies the critical junction length $J^\ast$ above which a wear debris is formed. This holds for both PH and HD formulations. Although based on simple observations, these calculations allow us to retrieve the transition curve in Eq.\,\eqref{eq:J-star}. They further permit to highlight the dependency of $J^\ast$ on the ratio $\ell/D$.

\subsection{Influence of the regularization parameter}
Calculations in Section \ref{sec:theory} lead us to extend our analysis to the three-dimensional space identified by the parameters $J$, $H/D$ and $\ell/D$. 
The aim is to understand whether the observed wear mechanisms change as a function of $\ell/D$.
Computations are conducted by using a HD variational formulation, for junctions with given asperity basis $D$, elastic moduli and fracture toughness.

Due to the extensive number of simulations needed to reproduce an entire chart $(J, H/D)$ for different ratios $\ell/D$, we restrict our attention to a representative selection of junction's geometries.
With reference to Figure \ref{fig:hydroD-chart}, the following two groups are considered.
\begin{itemize}
\item Geometries $\mathcal{G}_1$, with junction length $J=0.3$ and slenderness $H/D\in [0.3, 0.4, 0.5, 0.6, 0.7]$. For $\ell/D=0.02$, those junctions failed under the action of \textit{shear stresses}, through either a slip mechanism or the detachment of a small wear particle.
\item Geometries $\mathcal{G}_2$, with junction length $J=0.7$ and slenderness $H/D\in [0.2, 0.3, 0.4, 0.5]$. For $\ell/D=0.02$, failure was triggered by \textit{tensile stresses}, resulting in the formation of a small or large wear debris.
\end{itemize}

The ratio $\ell/D$ has to be small enough for the regularized phase-field formulation to $\Gamma$-converge to the sharp Griffith's model. We thus choose $\ell/D\in [0.005, 0.01, 0.015, 0.02, 0.025, 0.03]$, so that the characteristic size $D$ of the computational domain is at least $30$ times bigger than the regularization parameter $\ell$. For each value of $\ell/D$, the mesh size $\delta$ is chosen such that the numerical toughness $G_\text{c}^\text{num}=G_\text{c}(1=3\delta/8\ell)$ is the same in all the performed simulations.

\begin{figure}[hp!]
  \centering
  \subfloat[]
  {\includegraphics[width=0.75\linewidth]{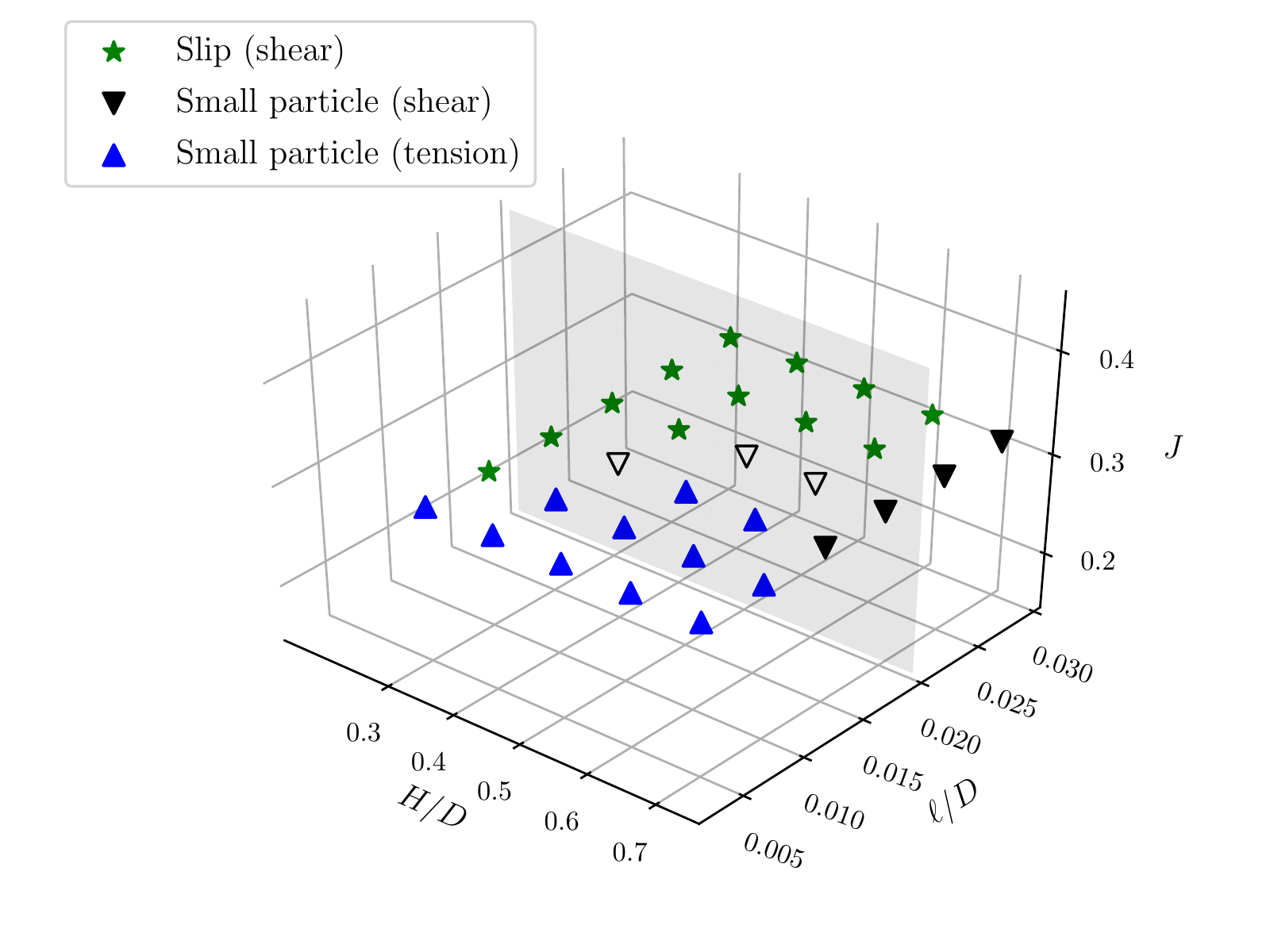}}\\
  \subfloat[]
    {\includegraphics[width=0.75\linewidth]{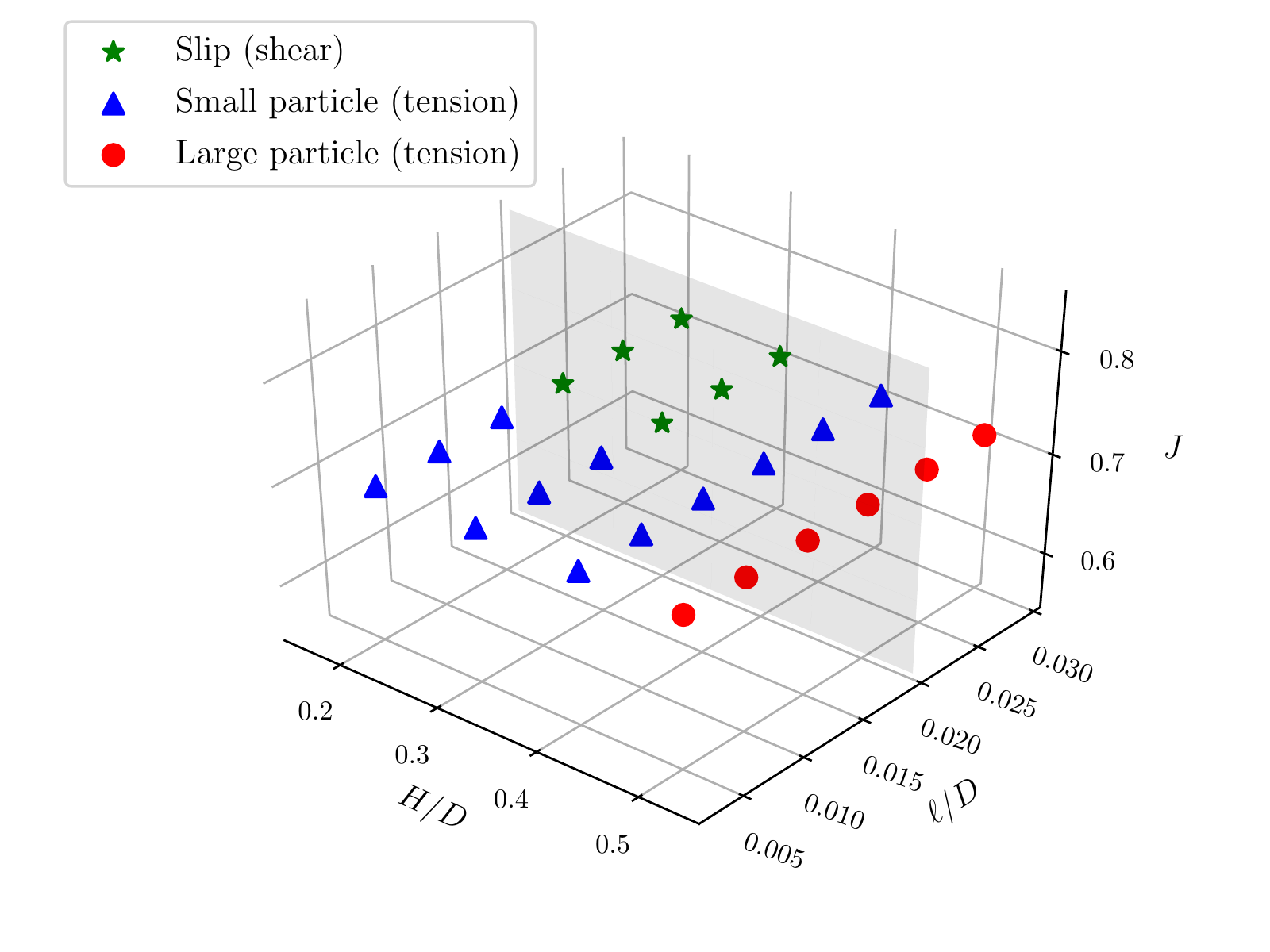}}
  \caption{
  Evolution of the failure mechanisms in the three-dimensional space defined by the junction's length $J$, the asperity's slenderness $H/D$ and the ratio $\ell/D$. HD conditions are assumed. (a) Geometries $\mathcal{G}_1$ with $J=0.3$ and slenderness $H/D\in [0.3, 0.4, 0.5, 0.6, 0.7]$. (b) Geometries $\mathcal{G}_2$ with $J=0.7$ and slenderness $H/D\in [0.2, 0.3, 0.4, 0.5]$. The regularization length $\ell$ of the phase-field model is varied such that $\ell/D\in [0.005, 0.01, 0.015, 0.02, 0.025, 0.03]$. The asperity basis is $D=1$. A grey shaded area indicates the plane $(J, H/D)$ corresponding to the ratio $\ell/D=0.02$, for which simulations in Figure \ref{fig:hydroD-chart} were conducted. 
  As in Figure \ref{fig:hydroD-chart}, the hollow markers represent the junctions that failed by developing one straight shear band, as shown in Figure \ref{fig:hydroD-mech} (b). 
  }
  \label{fig:3D}
\end{figure}

Obtained results are reported in Figure \ref{fig:3D}.
A grey shaded area indicates the plane $(J, H/D)$ corresponding to the ratio $\ell/D=0.02$, for which simulations in Figure \ref{fig:hydroD-chart} were conducted. With respect to those results, the following is observed.
\begin{itemize}
\item Geometries $\mathcal{G}_1$ in Figure \ref{fig:3D} (a). 
For large values of $\ell/D$, failure is mainly triggered by shear stresses. 
Depending on the slenderness $H/D$ of the asperity junction, two wear mechanisms are observed: small asperities (i.e., with relatively small values of $H/D$) slip one past the other, whereas  more slender junctions (i.e., with larger slenderness $H/D$) fail through nucleation of shear bands and detachment of a small wear particle.
As the ratio $\ell/D$ decreases, both those shear mechanisms are progressively replaced by tension-driven fracture.
For the smallest regularization length, small particle formation is observed for all the considered values of $H/D$.
\item Geometries $\mathcal{G}_2$ in Figure \ref{fig:3D} (b). Large ratios $\ell/D$ result in junctions failing by either slip, small or large particle formation, depending on the slenderness $H/D$ of the asperity. In particular, small junctions experience shear-driven failure, whereas more slender ones fail through the nucleation of tensile cracks. As the ratio $\ell/D$ decreases, the slip mechanism is replaced by small particle formation triggered by tensile stresses. On the other hand, no influence of $\ell/D$ is observed on junctions detaching a large particle debris.
\end{itemize}

Therefore, reducing the regularization length $\ell$ promotes tension-driven failure to the detriment of shear-driven mechanisms.
Interestingly, such a transition from slip to particle formation could only be observed via Molecular Dynamics \cite{brink-2019} by varying the strength $\widetilde{\tau}_\text{adh}$ of the adhesive interface with respect to the material bulk strength $\widetilde{\tau}_\text{b}$ (kept constant). 
According to this study, a junction that detaches a wear particle in presence of a fully adhesive interface ($\widetilde{\tau}_\text{adh}/\widetilde{\tau}_\text{b}=1$) can on the other hand fail through a slip mechanism when weakly-bonded ($\widetilde{\tau}_\text{adh}/\widetilde{\tau}_\text{b}<1$). 
 Our results further suggest that such a transition in failure mechanisms can also be experienced by perfectly-bonded adhesive junctions, as the material strength $\sigma_\text{c}$ varies with $\ell$.
 
 We further observe that, according to Eq.\,\eqref{eq:sigC}, the material strength $\sigma_\text{c}$ increases as the regularization length $\ell$ reduces. The maximum strain energy $\mathcal{E}_\text{el}$ in Eq.\,\eqref{eq:Eel} (that is, the energy stored in the junction prior to failure) increases correspondingly, while the fracture energy $\mathcal{E}_\text{fr}$ remains the same.
 This does not particularly affect junctions that fail through the detachment of a large particle, as they already meet the energy condition $\mathcal{E}_\text{el}\geq \mathcal{E}_\text{fr}$. Conversely, relatively small junctions (as those shown in Figure \ref{fig:3D}) are particularly sensitive to changes in the material strength, as an increase in $\sigma_\text{c}$ would allow them to store more energy in a rather small debris particle $A_\text{deb}$. Hence, as $\ell$ decreases, $\mathcal{E}_\text{el}$ might in fact overcome $\mathcal{E}_\text{fr}$. Indeed, this is what we observe in Figure \ref{fig:3D}  as the ratio $\ell/D$ reduces: for the smallest value, all the considered geometries fail by detaching a wear debris.

\begin{figure}[htb!]
  \centering
  \includegraphics[width=0.7\linewidth]{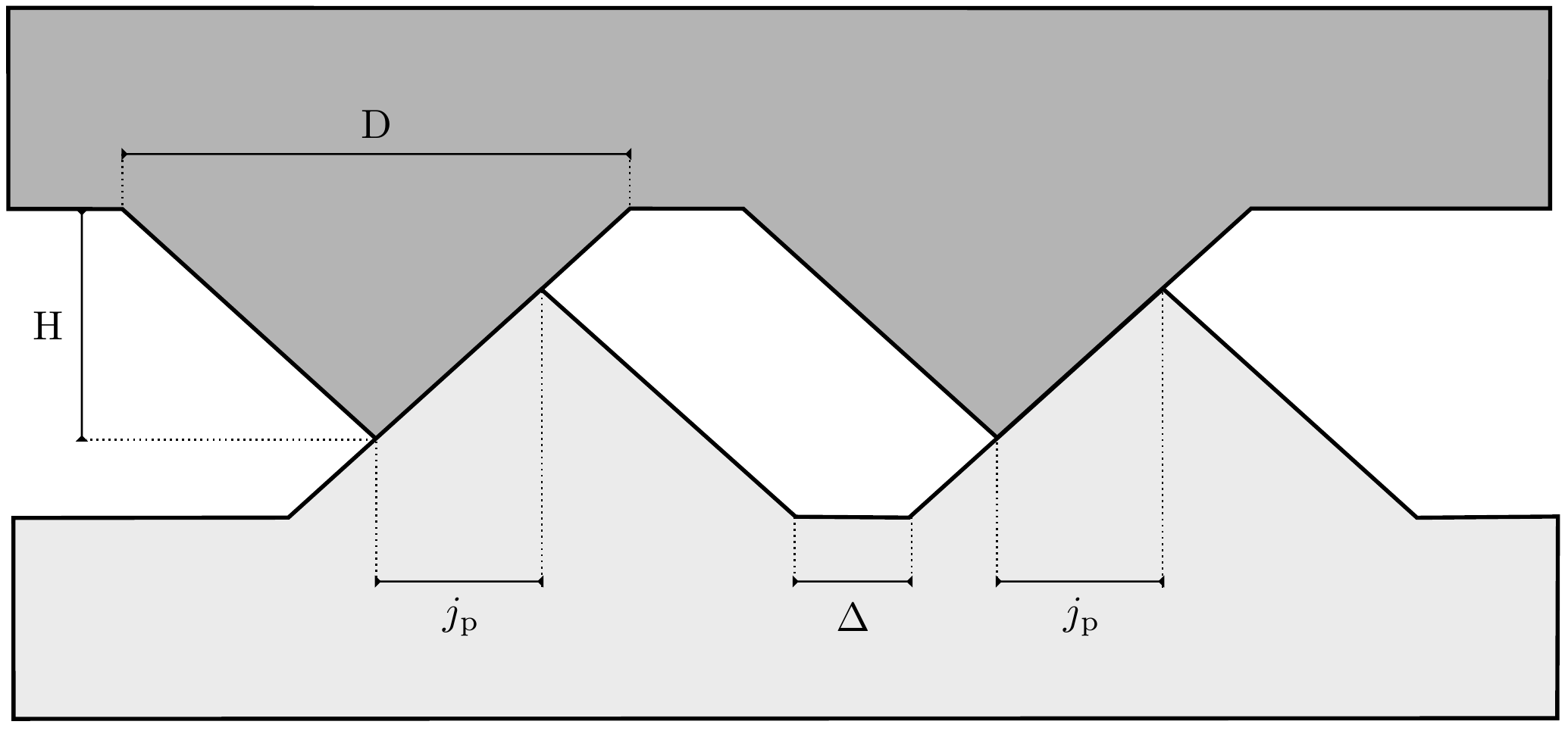}
  \caption{Schematic representation of the domain used for the study of interacting junctions. Asperities have the same slenderness $H/D$ and junction parameter $J$. The two junctions are are separated by an horizontal distance $\Delta$. The length $j_\text{p}$ is the horizontal projection of the junction.}
  \label{fig:inter-geom}
\end{figure}

\begin{figure*}[!ht]
  \centering
  \includegraphics[width=0.6\linewidth]{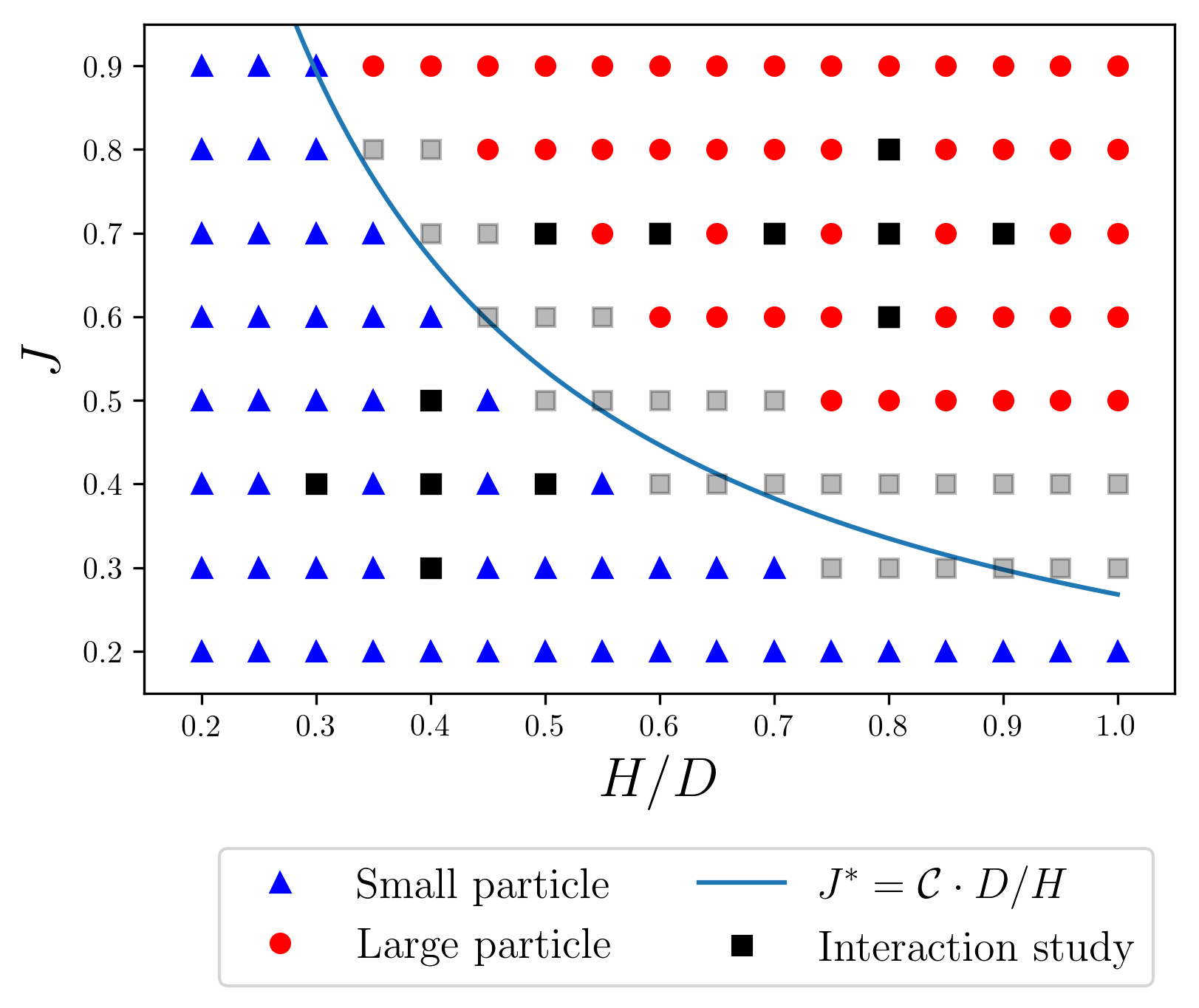}
  \caption{Representation of the geometries used to study the interaction between two junctions. The chart shows the failure mechanisms for single asperities under PH conditions (i.e., Figure \ref{fig:posH-chart}), where the rectangular markers highlight the geometries used for the interaction study. The black rectangles represent geometries far from the transition curve, which lead to the results represented as dense points in the interaction study in Figure \ref{fig:inter-chart}. The grey rectangles represent geometries sitting on each side of the transition curve, which lead to the results represented as transparent points in Figure \ref{fig:inter-chart}. $\mathcal{C} = 0.27$.}
  \label{fig:inter-single-chart}
\end{figure*}

\begin{figure}[hbt!]
 \centering
  \begin{overpic}[width=0.45\textwidth]{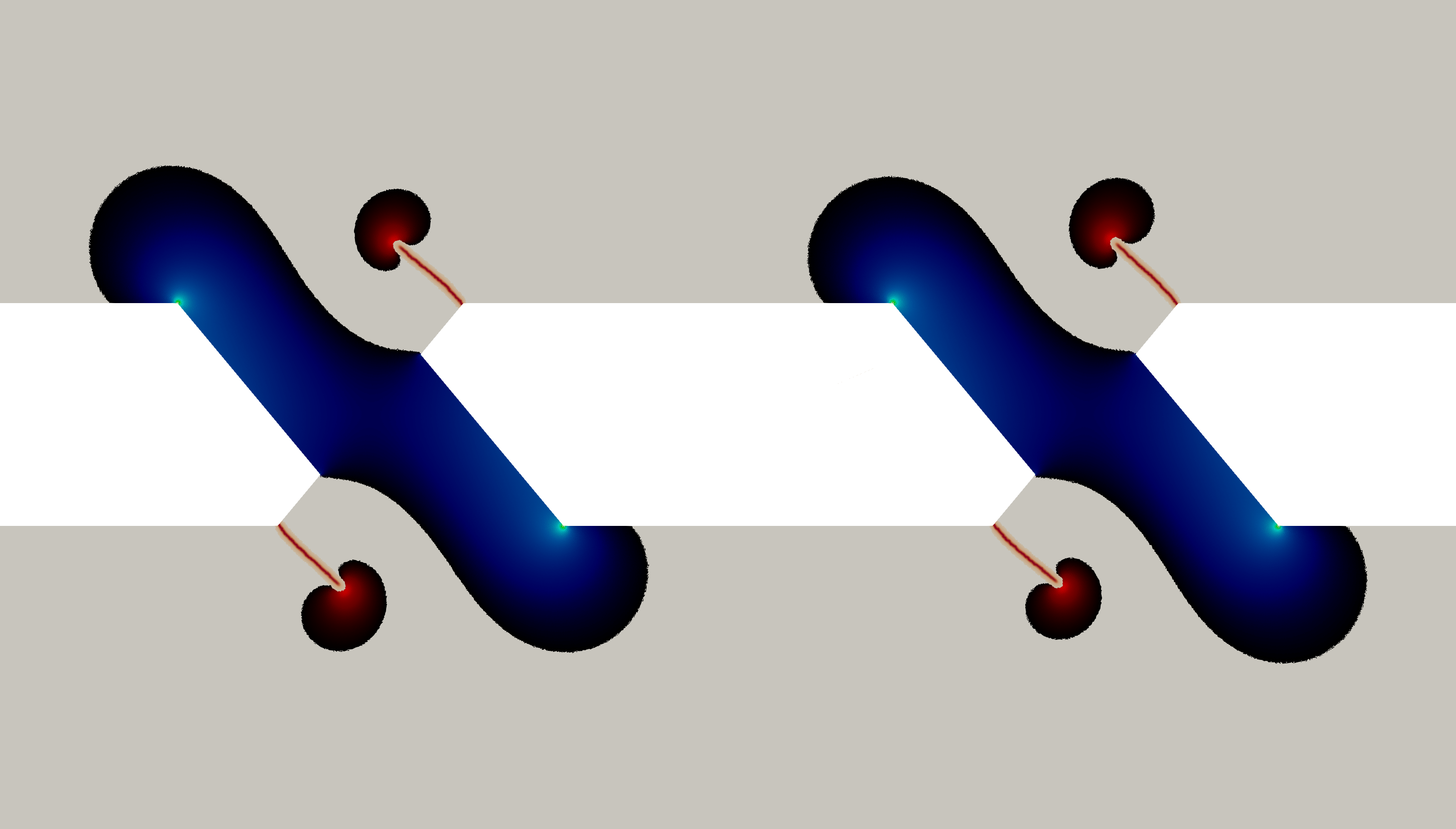}
    \put (5,50) {(a)}
  \end{overpic} \hspace{1cm}%
  \begin{overpic}[width=0.45\textwidth]{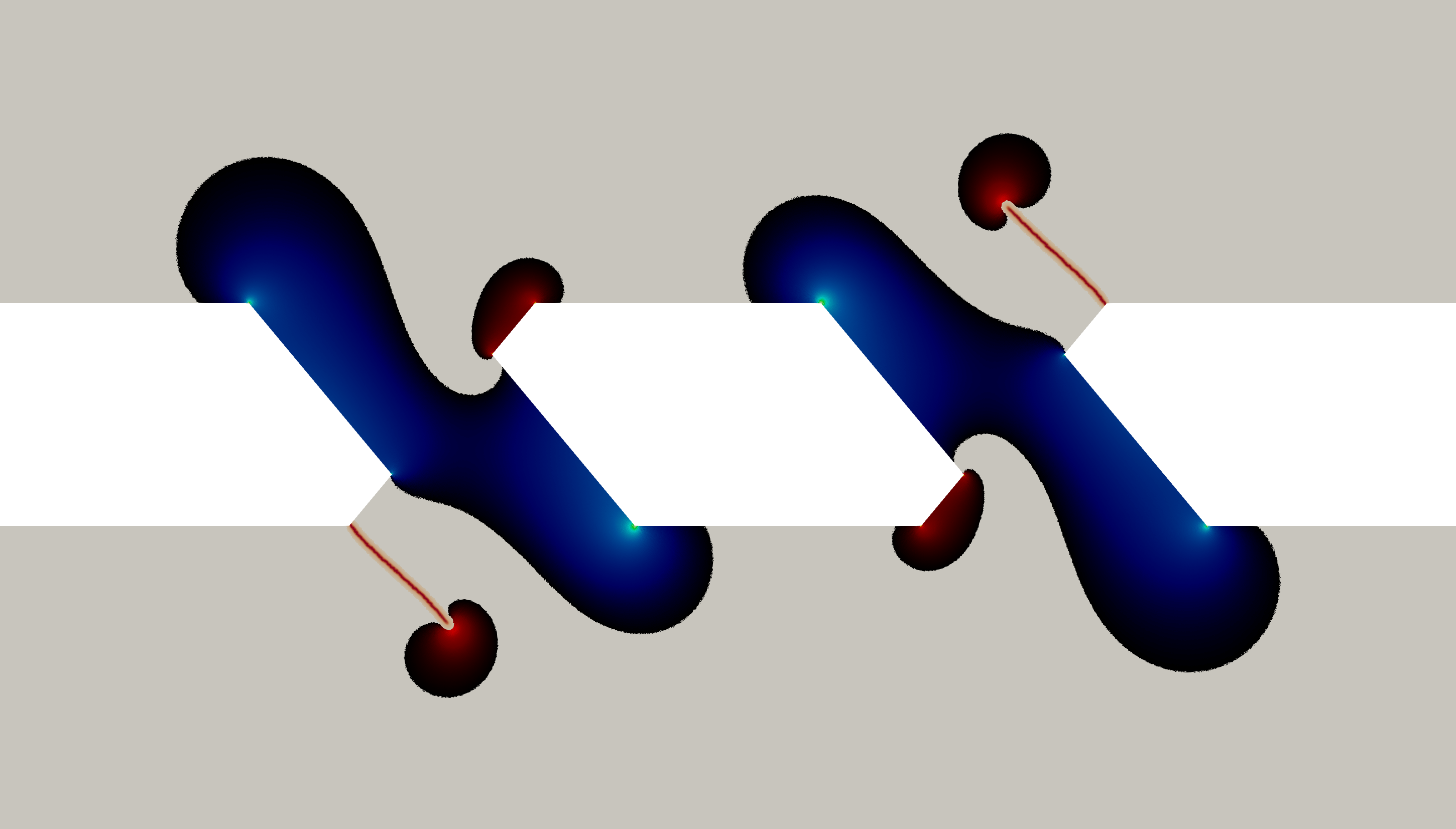}
    \put (5,50) {(b)}
  \end{overpic}
  \caption{Interaction study under PH conditions, for a cluster of junctions with $J=0.7$ and $H/D=0.6$. Effects of the distance between two junctions on the observed failure mechanisms. The hydrostatic stresses and damage fields are displayed according to the thresholds in Figure \ref{fig:mesh-length}.  (a) The two junctions are separated by an horizontal distance $\Delta=1.5D$. A pair of cracks independently nucleates in the bulk around each junction. The resulting wear product is identified as large debris formation. (b) The same junctions as in (a), where the horizontal distance is now reduced to $\Delta=D$. Only one pair of cracks nucleates and the two junctions will eventually detach as a single wear debris. This phenomenon is identified as interacting junctions or macro-particle formation. Note that the figures show zoom-ins, cropped from a much bigger domain.}
  \label{fig:inter-mech}
\end{figure}

\section{Interactions among adhesive junctions}
\label{sec:interactions}
As briefly recalled in the Introduction, the most extreme wear rates observed might be associated with the detachment of macro-particles whose relevant length-scale exceeds a single junction \cite{aghababaei-2018}. Thus, any realistic approximation of the wear rate must account for the presence of clusters of junctions rather than considering each of them as an isolated system. We therefore study how two junctions interact with each other when getting closer together. The simulations follow the same procedure as described in Section \ref{sec:mesh-load-cond}, although the computational  domain now consists of two junctions as represented in Figure \ref{fig:inter-geom}. The regularization length is such that $\ell/D=0.02$.

We use a similar approach as found in \cite{pham-2019} which defines a \textit{real} ($j_\text{r}$) and an \textit{apparent} ($j_\text{a}$) contact length. The former is the actual contact length developed within the junction, while the latter is the total length of the cluster which also includes the space $\Delta$ between the asperities. We use the horizontal projection $j_\text{p} =J D/2$ to express the contact length. In order to compare various junction's geometries, we normalize both $j_\text{r}$ and $j_\text{a}$ with the projected critical junction length
\begin{equation}
  \label{eq:jpstar}
  j^*_\text{p} = j_\text{p} \, J^\ast = \frac{D^2}{2H} \, \mathcal{C}
\end{equation}
which represents the junction length at which a single junction with a slenderness $H/D$ shows a transition from small to large particle formation. This normalization parameter is computed from the transition curve identified in the study of single asperities. The real junction length is then defined as the normalized sum of the projected junction lengths $j_\text{p}$ (see Figure \ref{fig:inter-geom})
\begin{equation}
  \label{eq:jr}
  j_\text{r} = \frac{j_\text{p}}{j^*_\text{p}}
\end{equation}
which corresponds to a system with two junctions. The apparent contact length is defined as the normalized sum of the asperity length and the distance $\Delta$

\begin{equation}
  \label{eq:ja}
  j_\text{a} = \frac{1}{j^*_\text{p}}\left( D + \frac{\Delta}{2}\right)
\end{equation}

\begin{figure*}[htb!]
  \centering
  \includegraphics[width=0.65\linewidth]{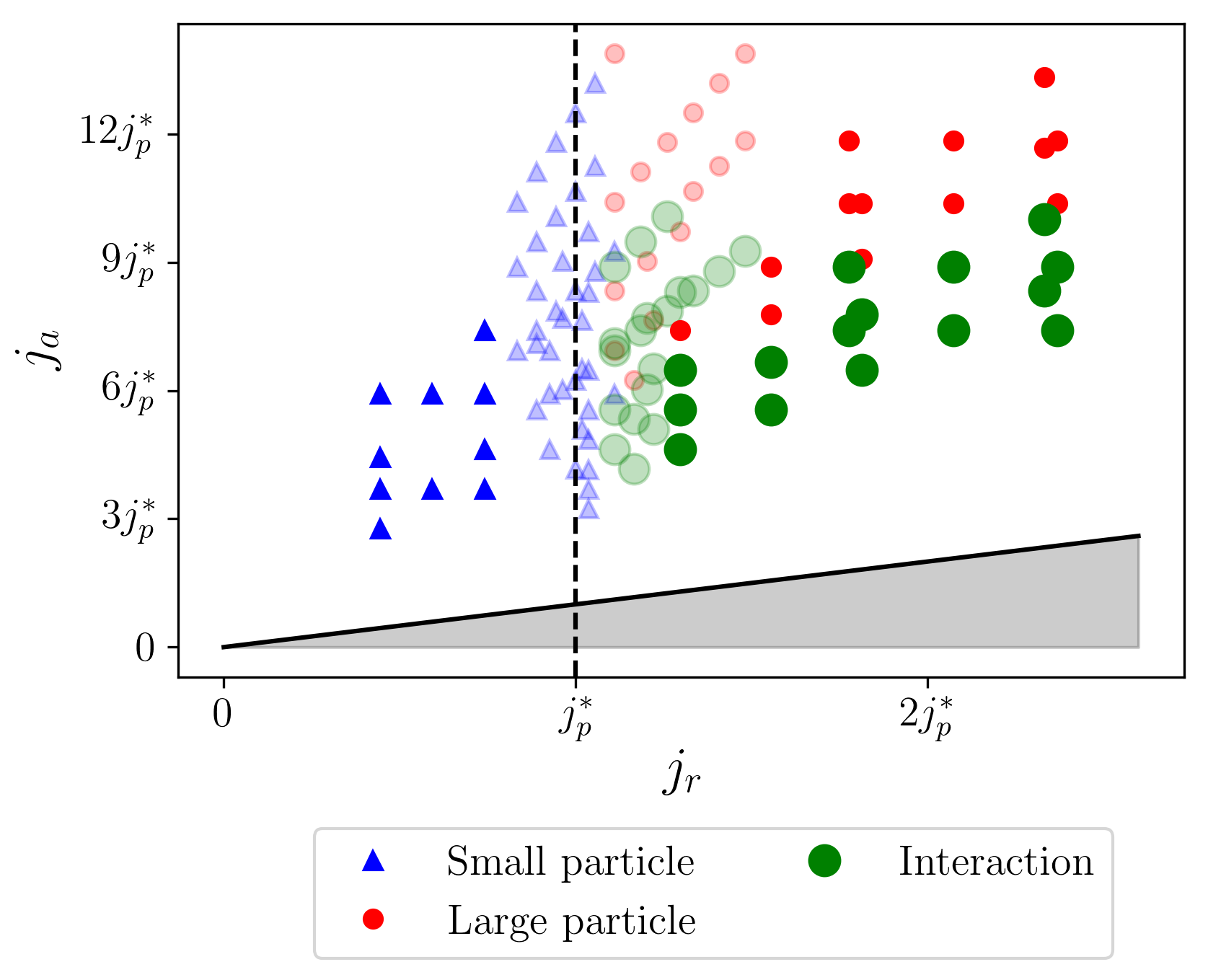}
  \caption{
  Interaction study, evolution of the failure mechanisms as a function of the real $j_\text{r}$ (Eq.\,\ref{eq:jr}) and apparent $j_\text{a}$ (Eq.\,\ref{eq:ja}) contact lengths. The blue triangles represent independent failures of the two junctions through the nucleation of cracks inside the junction, resulting in small particle formation. The red dots represent independent failures of the two junctions through the nucleation of cracks inside the bulk (see Figure \ref{fig:inter-mech} (a)), resulting in large particle formation. The large green dots represent interacting failures of the junctions through the nucleation of a single pair of cracks that embodies the two junctions (see Figure \ref{fig:inter-mech} (b)), resulting in the formation of a macro-particle debris. The dense markers are the results of interaction studies performed on geometries represented by dense markers in Figure \ref{fig:inter-single-chart}, which lie far from the transition curve. Transparent markers correspond to junction geometries represented by transparent points in Figure \ref{fig:inter-single-chart}, which lie on either side of the transition curve. }
  \label{fig:inter-chart}
\end{figure*}

When considering the formation of a macro-particle that encapsulates several asperities, it is natural to expect the cracks to form in the bulk rather than in the junction. It is indeed very unlikely that junctions that independently fail with a mechanism where the crack grows only within the junction would interact to form a macro-particle.
We thus expect that the macro-particle formation would only be located above the transition curve shown on Figures \ref{fig:posH-chart} and \ref{fig:hydroD-chart} (red dots). The comparative study of PH and HD models shows that this region is independent of the chosen formulation.
We thus perform the interaction study by referring to the PH formulation.\footnote{Here we make the assumption that a macro-particle can only be formed from two junctions which individually fail under tension (that is, above the transition curve) while the geometries that fail under shear (below the transition curve) are assumed to never interact. Indeed, a comparative study under HD conditions is needed to assess the role played by shear stresses and will be the object of future investigation.}

As the failure mechanism strongly depends on the geometry, we consider junctions that correspond to various modes of failure to conduct a thorough study of the interaction. The rectangular markers in Figure \ref{fig:inter-single-chart} show the $\{H/D,J\}$ couples used to study the failure of a domain with two junctions. For each of these geometries, we perform several simulations by varying the distance $\Delta$ between the asperities, which ultimately allows us to express the evolution of the failure mechanisms as a function of the real ($j_\text{r}$) and apparent ($j_\text{a}$) contact lengths. The obtained results are shown in Figure \ref{fig:inter-chart}.

We observe that geometries that lie far from the transition curve in Figure \ref{fig:inter-single-chart} (black rectangular markers), give rise to very well defined regions in Figure \ref{fig:inter-chart} (dense markers). On the other hand, the geometries located close to the transition curve in Figure \ref{fig:inter-single-chart} (grey rectangular markers) tend to produce results that slightly overlap in Figure \ref{fig:inter-chart} (transparent markers). Regardless of the distance $\Delta$, every junction that independently fails forming a small particle shows the same behavior when considered as a group. In Figure \ref{fig:inter-chart}, this corresponds to the area identified by small real contact lengths $j_\text{r}$  (blue triangles). Conversely, two distinct regions are observed for larger values of $j_\text{r}$. Geometries with a high $j_\text{a}$ exhibit a crack pattern similar to that shown in Figure \ref{fig:inter-mech} (a) (red dots). These points correspond to junction couples where the distance $\Delta$ is large enough to prevent interaction, resulting in the detachment of a large particle debris. Geometries with a lower $j_\text{a}$ show a crack pattern similar to that in Figure \ref{fig:inter-mech} (b) (large green dots). In these cases, the distance $\Delta$ is small enough for the interaction to occur and the two junctions detach together as a macro-particle.

\section{Discussion and concluding remarks}
\label{sec:discussion}
This paper investigates adhesive wear in asperity junctions by using a variational phase-field approach to brittle fracture \cite{francfort-1998, bourdin-2000, bourdin-2008}. We consider two regularized formulations with contact conditions, as proposed in \cite{amor-2009}.
In the positive-hydrostatic model (PH), fracture is only allowed in material regions with positive volume change, while the hydrostatic-deviatoric fomulation (HD) also includes shear-driven fracture.

\begin{figure*}[b!]
  \centering
    \begin{overpic}[width=0.7 \linewidth]{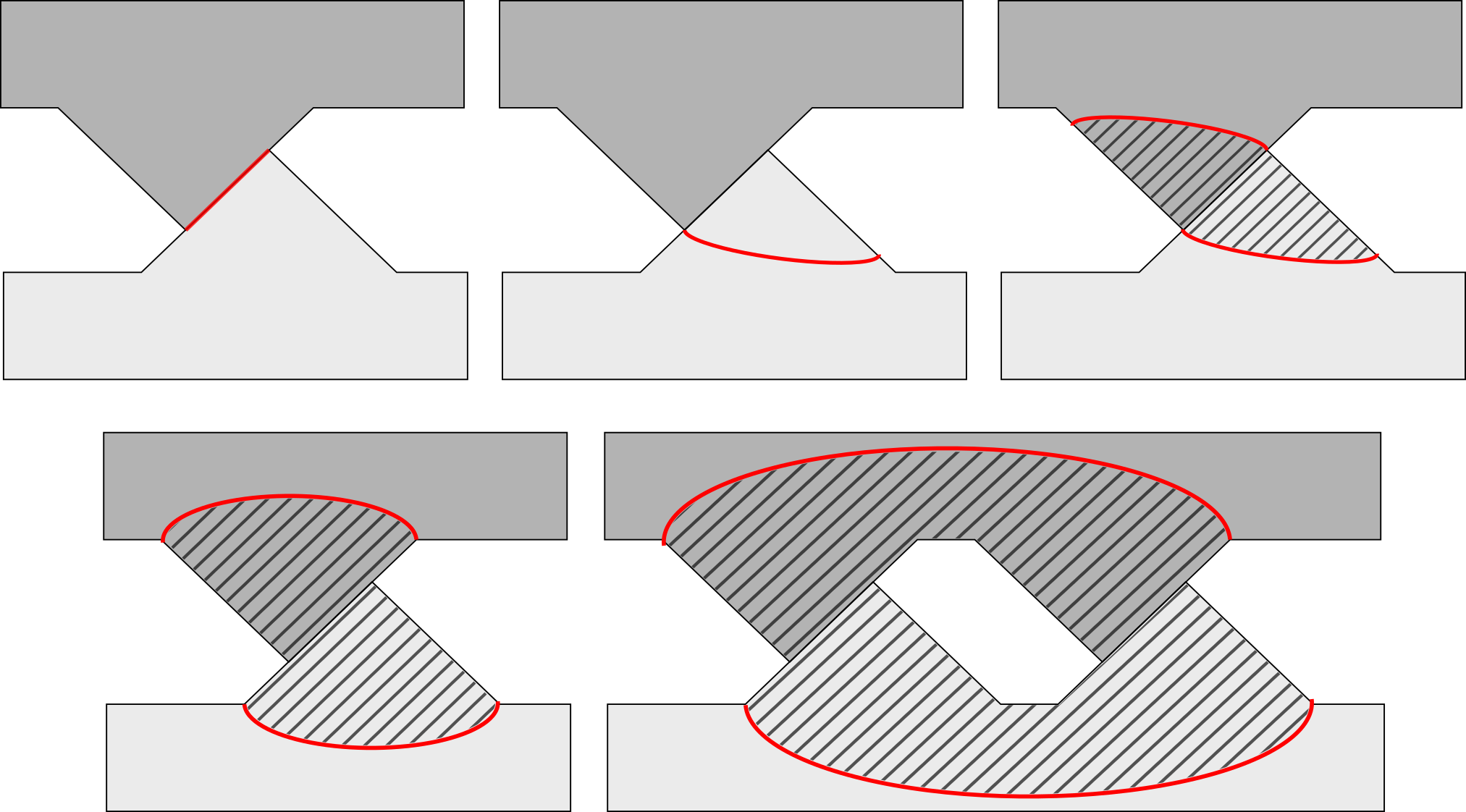}
        \put (1,52) {(a)}
        \put (35,52) {(b)}
        \put (69,52) {(c)}
        \put (7,22) {(d)}
        \put (42,22) {(e)}
    \end{overpic}
    \caption{Schematic representation of the observed failure mechanisms. Red curves and shaded areas respectively represent cracks and the produced wear debris.
    (a)  Slip: a shear band propagates along the interface between the asperities; no debris formation occurs. (b) Single crack in the junction: a single shear band propagates straight from one of the stress concentrations in the junction; no debris formation occurs. (c) Small particle formation: a pair of cracks propagates in the junction; small debris formation occurs. (d) Large particle formation: a pair of cracks propagates in the bulk; large debris formation occurs. (e) Macro-particle formation: two junctions interact and a pair of cracks propagates in the bulk.}
    \label{fig:wear-form-single}
\end{figure*}
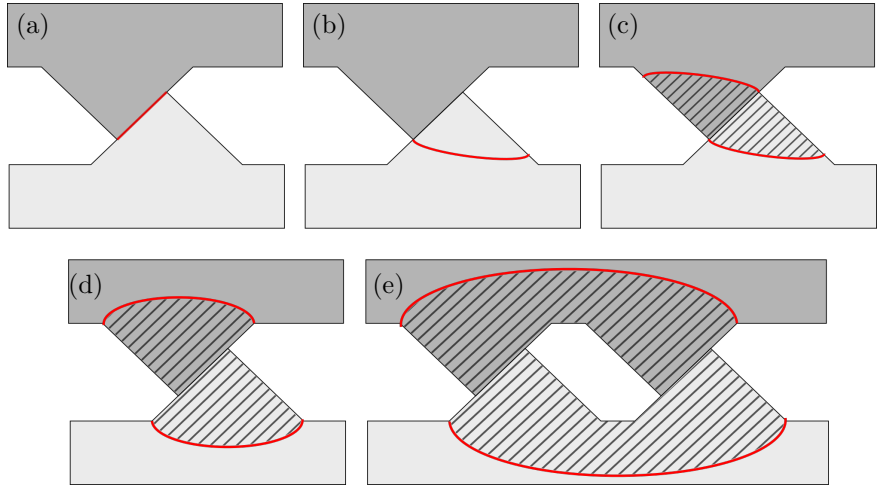

We first consider the case of a single junction comprised of two triangular asperities in contact to each other. 
The crack paths observed in the comparative study of various geometries lead to the identification of various failure patterns. This gives rise to a classification in distinct failure mechanisms (schematically summarized in Figure \ref{fig:wear-form-single} (a)--(d)), each of them is characterized by a triggering stress (shear or tension) and a debris formation pattern (no particle, small particle and large particle detachment).

The study of single junctions under PH conditions shows that tension-driven cracks results in the formation of either a \textit{small} (Figure \ref{fig:wear-form-single} (c)) or \textit{large particle} (Figure \ref{fig:wear-form-single} (d)), depending on the slenderness $H/D$ and on the relative overlap $J$ of the asperities forming the junction. 
We find that the transition from small to large particle formation can be described by the criterion $J^*=D/H\cdot \mathcal{C}$ (transition curve), where $\mathcal{C}$ is a constant parameter depending on the material properties. 
With respect to the failure mechanisms observed under PH conditions, the HD formulation allows us to capture the occurrence of two additional failure modes, \textit{slip} and \textit{shear}. When slip occurs, a shear band forms along the interface between the asperities and no debris is produced (Figure \ref{fig:wear-form-single} (a)). On the other hand, when a shear mode is observed, a horizontal shear band develops within one or both asperities and the junction fails with either no debris formation (Figure \ref{fig:wear-form-single} (b)) or by releasing a small particle (Figure \ref{fig:wear-form-single} (c)). The criterion $J^*=D/H\cdot \mathcal{C}$, expressed in terms of the same material constant $\mathcal{C}$,  now describes the change in the stresses triggering failure from tension ($J>J^*$) to shear ($J<J^*$). Interestingly, the condition for large particle formation ($J>J^*$) is independent from the chosen coupling between elastic energy and damage field. These results are in agreement with observations reported in Molecular Dynamics studies \cite{aghababaei-2016} highlighting the existence of a critical length scale, which uniquely identifies the transition from relatively low mass loss to large particle detachment.
Such a condition for particle formation is analytically derived based on the arguments presented in \cite{aghababaei-2016}. This leads us to investigate the influence of the phase-field regularization parameter on the observed failure mechanisms. Results show that reducing the regularization length (that is, increasing the material strength) promotes tension-driven failure to the detriment of shear-driven mechanisms, as a transition from slip to particle formation is observed.

In the second part of the paper, we study the interaction between two asperity junctions and show how microcontact interactions can be favoured in some geometries to form macro-particles (Figure \ref{fig:wear-form-single} (e)). We perform the interaction study by using a PH formulation, as we expect macro-particle detachment from geometries which independently produce a large debris ($J>J^*$). 
We find that the two junctions start interacting as the distance between them decreases, eventually detaching as a single wear debris or macro-particle. The transition from a cluster of non-interacting junctions releasing a large debris to a cluster of interacting junctions producing a macro-particle is clearly identified in terms of real and apparent contact lengths. 

Finally, a central point in our analysis consists in the relation between the debris formation patterns and the macroscopic wear rate induced by the failure of the adhesive junctions.
Since mild and severe wear regimes are usually associated to particle detachment through crack propagation, we expect our phase-field approach to be  effective in the prediction of those wear processes. The same logic would however put the low wear regime out of our reach, since this wear form is commonly associated to the complete alteration of the domain through large deformations. 
Although serving as a first step and approximation for more refined calculations, the proposed results do deliver some useful insights on the low wear regime, as the failure mechanisms (a) and (b) in Figure \ref{fig:wear-form-single} are driven by shear and do not lead to any particle formation, which indeed are two features usually associated to this wear form.

Based on the discussion above, we propose a classification in terms of wear rate of the five failure mechanisms shown in Figure \ref{fig:wear-form-single}.
\begin{itemize}
\item The mechanisms that do not form any wear debris (Figure \ref{fig:wear-form-single} (a) and (b)), which are driven by shear,  can be associated to a \textbf{low wear} regime.
\item The mechanisms that form small or large wear debris (Figure \ref{fig:wear-form-single} (c) and (d)), which are driven by tension, can be associated to a \textbf{mild wear} regime.
\item The mechanisms that form macro-particles (Figure \ref{fig:wear-form-single} (e)), which are driven by tension, can be associated to a \textbf{severe wear} regime.
\end{itemize}
This change in the wear process at the microscopic level could explain the change in the type of relation between wear rate and applied load at the macroscopic one. It is however to be noted that, in a large-scale friction process, several failure mechanisms can simultaneously contribute to the overall wear, in which case probabilistic approaches are more suitable to estimate the macroscopic wear rate \cite{frerot-2018}.

To summarize, we were able to recover clear transitions in the crack nucleation patterns using a phase-field variational approach to fracture. Based on the results from Molecular Dynamics simulations found in literature, we proposed to associate each crack nucleation pattern to a failure mechanism. The geometry of the adhesive junctions is understood to be responsible for the change in failure mechanism, which can be expressed as a function of the junction length and the slenderness of the asperities. This transition curve could be recovered regardless of the coupling between the damage and the elastic energy. For geometries located above the transition curve, the spacing between the junctions determines if they interact to detach as a single macro-debris. Prescribing different coupling between the damage field and the elastic energy allowed us to assess the triggering mechanisms underlying the various failure processes: large and macro-particle formation is associated with tension-driven cracks while small or no particle formation is typically associated with the formation of shear bands. With this distinction, we could recover all failure mechanisms associated to adhesive junctions found in the literature: slip, gradual smoothing, small, large and macro-debris formation.

Finally, this study showed the potential of variational phase-field formulations in studying adhesive wear in brittle materials. Moving forward, future work shall investigate the tribological behavior of elastic-plastic junctions, which on the other hand is hardly accessible by discrete numerical models such as those based on Molecular Dynamics.

\paragraph*{Acknowledgment}
The authors thank the Zeno Karl Schindler foundation for its financial support and Prof. Kaushik Bhattacharya for providing a working space as well as computational resources at the California Institute of Technology.

\newpage
\bibliography{sc-pdm.bib}

\end{document}